\documentclass[12pt,preprint]{aastex}

\def\kms{\mbox{${\rm km}\:{\rm s}^{-1}\,$}}
\def\kmsb{\mbox{${\rm km}\:{\rm s}^{-1}$}}

\newcommand{\Teff}{$T_{\!\mbox{\scriptsize\em eff}}$}

\newcommand{\hii}{H{\sc ii}\rm}

\newcommand{\fglr}{\sc fglr~\rm}

\newcommand{\bb}{\ensuremath{\mathrm{\AA}}} 

\slugcomment{}

\shorttitle{Distance and Metallicity for WLM}
\shortauthors{Urbaneja et al.}

\begin{document}


\title{The Araucaria Project: The Local Group Galaxy WLM - Distance and Metallicity from Quantitative Spectroscopy of Blue Supergiants\altaffilmark{1}}

\author{Miguel A.~Urbaneja} \affil{Institute for Astronomy, 2680 Woodlawn
Drive, Honolulu, HI 96822; urbaneja@ifa.hawaii.edu}

\author{Rolf-Peter Kudritzki} \affil{Institute for Astronomy, 2680 Woodlawn
Drive, Honolulu, HI 96822; kud@ifa.hawaii.edu}

\author{Fabio Bresolin} \affil{Institute for Astronomy, 2680 Woodlawn
Drive, Honolulu, HI 96822; bresolin@ifa.hawaii.edu}

\author{Norbert Przybilla} \affil{Dr. Remeis-Sternwarte Bamberg,
Sternwartstr. 7, D-96049 Bamberg, Germany;
przybilla@sternwarte.uni-erlangen.de}

\author{Wolfgang Gieren} \affil{Universidad de Concepci\'on, Departamento de Astronom\'{\i}a, 
Casilla 160-C, Concepci\'on, Chile; wgieren@astro-udec.udec.cl}

\and

\author{Grzegorz Pietrzy\'nski} \affil{Universidad de Concepci\'on, Departamento de Astronom\'{\i}a, 
Casilla 160-C, Concepci\'on, Chile;  pietrzyn@hubble.cfm.udec.cl}

\altaffiltext{1}{Based on VLT observations for ESO Large Programme 171.D-0004.}

\begin{abstract}

The quantitative analysis of low resolution spectra of A and B 
supergiants is used to determine a distance modulus of 24.99$\,\pm\,$0.10 mag 
(995$\,\pm\,$46 Kpc) to the Local Group galaxy WLM. The analysis yields 
stellar effective temperatures and gravities, which provide a distance through the 
Flux weighted Gravity--Luminosity Relationship ({\sc fglr}). Our distance is 0.07~mag 
larger than the most recent results based on Cepheids and the tip of the RGB. This difference is 
within the 1$\displaystyle\sigma$ overlap of the typical uncertainties quoted in these photometric 
investigations. In addition, non-LTE spectral synthesis of the rich metal line spectra  
(mostly iron, chromium and titanium) of the A supergiants is carried out, which allows the 
determination of  
stellar metallicities. An average metallicity of -0.87$\,\pm\,$0.06~dex with respect to solar
metallicity is found.

\end{abstract}

 

\section{Introduction}\label{intro}


WLM is one of the faintest dwarf irregular galaxies ($M_{B} \cong -14$), located 
in an isolated part of the Local Group. Detailed photometric studies have shown that
it consists of a young population concentrated in a disk and an old extended 
metal poor halo \citep{ferraro1989,minniti1997,dolphin00,rejkuba2000,mcconnachie2005}. 
The color of the red giant branch of 
the old population indicates a metallicity of [Fe/H] = -1.45 dex\footnote{We 
use the common notation [X/Y] = log(X/Y) - log(X/Y)$_\odot$, with the solar
abundances given by \cite{grevesse98}, except for oxygen, for which we adopt the value from 
\cite{asplund2004}.} representing the 
end of the first star formation episode. On the other hand, the metallicity of 
the young population is somewhat ambiguous. While measurements from a number of
\hii~regions \citep{skillman1989,hodge1995,lee2005} yielded a low 
nebular oxygen abundance of [O/H]\,=\,-0.8\,dex, a detailed high resolution 
spectroscopic study of an A-type supergiant \citep{venn2003} obtained 
[O/H]\,=\,-0.2\,dex indicating a large discrepancy between 
stellar and nebular chemical composition. On the other hand, \citet{bresolin2006}
in their low resolution (5\,\AA) spectroscopic survey for supergiants in WLM 
studied three early B supergiants and found an average value of [O/H]\,=\,-0.8\,dex 
comparable with the \hii~regions.

The spectroscopic sample gathered in the survey by \citet{bresolin2006} also 
contained high signal-to-noise data of several late B and A supergiants 
(B5-A7) of luminosity class Ia and II. These objects with their rich spectra of
metal lines (mostly iron, titanium, chromium) are ideal 
for a further investigation of stellar metallicity. However, at the time of the
publication of this work the density of metal lines in the optical spectra 
together with the low spectral resolution did not allow for a determination of 
stellar parameters using the standard analysis techniques applied for the 
early-type B supergiants. Recently, this situation has changed.  
\citet[][hereafter K08]{kudritzki2008} in their study of the Sculptor spiral 
galaxy NGC\,300 at a distance of 1.9 Mpc developed a new technique to quantitatively 
analyze low resolution spectra of A supergiants with respect to stellar parameters 
and metallicity. This technique makes use of the Balmer jump and the Balmer lines to obtain 
effective temperature and gravity and determines metallicity from a 
$\chi^{2}$-minimization of observed and theoretical spectra in selected spectral windows. 
It can now be applied to the data set obtained by \citet{bresolin2006} and allows 
for a more comprehensive straightforward spectroscopic study of the metallicity
of the young stellar population in WLM.

The distance to WLM has been determined from photometric 
studies of the old stellar population using the tip of the red giant branch 
(TRGB), the horizontal branch (HB) and full color-magnitude diagrams, yielding  
distance moduli between 24.7 and 24.95~mag \citep[see][and references above]{rizzi2007}.
The most recent comprehensive multi-color survey for Cepheids by \citet{gieren2008}, including
J- and K-band photometry, found a distance modulus of 24.92~mag with a very small random error of 
0.04~mag and a systematic error of 0.05~mag. These authors determined an average interstellar 
reddening E(B-V)=0.08~mag, significantly larger than the Galactic foreground 
reddening and, thus, relevant for the derived distance. Under such circumstances, an 
independent distance and reddening determination using the young stellar population is  
ideal for verifying photometrically based distances.

Such an independent distance determination can be provided through the 
quantitative spectral analysis of the B and A supergiants. \citet{kudritzki2003}
and K08 have shown that a tight correlation exists between the 
absolute bolometric magnitude and the flux weighted gravity $g/T_{eff}^{4}$.
This relationship is predicted by stellar evolution theory and  
allows distance determinations with a precision comparable to Cepheids.

In this paper, we will carry out a spectral analysis of the low resolution 
optical spectra of 8 late B and A supergiants (B5-A7) observed by 
\citet{bresolin2006} to determine stellar temperatures, gravities, luminosities, 
masses and metallicities. For the analysis we will use the basic concept 
introduced by K08, however, in a modified form. We will present  
a new numerical fit algorithm based on an empirical application of the Karhunen-Lo\`eve expansion
usually referred to as Principal Component Analysis, which 
allows for an automated analysis of a large number of objects using large 
comprehensive grids of model atmosphere spectra. We will then combine the 
results with those obtained by \citet{bresolin2006} for the early-type 
B supergiants and determine an independent distance using the flux-weighted 
gravity luminosity relationship, {\sc fglr}. This will be the first distance determination 
using this new distance determination method. 

The paper is organized as follows. After a brief description of the 
observations in \S\ref{sec_obs} we present the modified analysis method in 
\S\ref{sec_anal}. In \S\ref{sec_hires}, we test our method by comparing with the 
results obtained in a previous detailed high resolution study for one of our targets. 
The results of the analysis, i.e. the derived \Teff, $log~g$, metallicity,
E(B-V), radius, luminosity and mass, are given in \S\ref{sec_results}. The 
distance determination using the \fglr is carried out in \S\ref{sec_fglr} 
and \S\ref{sec_discussion} will present a discussion of all the results.

\section{Observations}\label{sec_obs}

The spectroscopic observations were obtained with the Focal Reducer and Low Dispersion 
Spectrograph 2 \citep[FORS2,][]{appenzeller1998} at the ESO Very Large Telescope in 
multi-object spectroscopy mode in one night of good seeing conditions (better than 0.7 arcsec) 
on 2003 July 28 as part of the Auracaria 
Project \citep{gieren2005}. 
The total exposure time is 4500s and the air mass during the observations was 
smaller than 1.07. The spectral resolution is approximately 5~\AA~and the 
spectral coverage extends over 2500~\AA~centered at 4500~\AA~for most of our 
targets. The spectra were also flux calibrated so that spectral energy 
distributions, in particular the Balmer jump, can be used to constrain the 
stellar parameters. The observational data set has been described by 
\citet{bresolin2006}. The paper also contains finding charts, coordinates, 
photometry \citep[V, I from the Las Campanas 1.3m telescope, see][]{pietrzynski2007}, 
radial velocities and 
spectral types. Of the 19 confirmed supergiants found with spectral types ranging 
from late O to G, 8 were of spectral type B5 to A7. The S/N per pixel for these 
objects is between 40 to 120, sufficient for our quantitative analysis. This  
sample listed in Table~\ref{tab_objects} and selected for this work. The target 
designation follows the nomenclature in \citet{bresolin2006}. Note that 
we consider only stars of their A set because of the significantly better S/N.

Along with the broad band photometry in the aforementioned reference, we also 
use  B-, V-, R- and I-band data published by \citet{massey2007} to constrain 
the interstellar reddening E(B-V), as well as J- and K-band 
photometry from \citet{gieren2008}. The IR photometry, available only for some
stars, is compiled in Table~\ref{tab_IR}.

Prior to the quantitative spectral analysis, we carefully explore the dataset 
available for each star, to identify
spectral regions showing signs of contamination due to imperfect sky subtraction,
nearby cosmic rays, and similar other observational effects. These regions are 
masked out and not used in the analysis. 

\section{Quantitative spectral analysis \label{sec_anal}}

As explained in K08 the principal difficulty in the analysis of low resolution 
spectra of A supergiants lies in the simultaneous determination of effective 
temperature and metallicity. The classic method based on the use of ionization 
equilibria does not work, since the required lines from the neutral species 
(\ion{Mg}{1}, \ion{Fe}{1}, \ldots) are in general very weak and 
disappear at low resolution within noise and the blends of spectral lines. 
While one could use the information from the stronger lines, which define 
the spectral type, and a calibration of spectral type with effective 
temperature, such a relationship is metallicity dependent, which requires 
the simultaneous and independent determination of metallicity.

K08 were able to solve this problem by using the Balmer jump as an 
independent temperature indicator in addition to the Balmer lines, which can 
be used as a measure of gravity, and to the rich metal line spectrum, which 
yielded the metallicity through a $\chi^2$-minimization comparing 
selected windows of the observed spectra with very sophisticated non-LTE line 
formation calculations.

The methodology employed in the present work follows closely the ideas 
presented by K08, however with some significant modifications. While K08 
constructed fit diagrams of the Balmer jump and the Balmer lines in the 
$T_{eff}$--$log~g$ plane through ``by eye'' fits of the observed line 
profiles and the spectral energy distribution (SED) and then determined the 
metallicity for the effective 
temperature and gravity at the intersection of the two fit curves. Here we have 
developed a completely automated numerical method, which finds the best fit of 
the model atmosphere synthetic spectra to the observed spectra. The advantages 
of this new method are obvious. It provides a more quantitative and objective 
way to obtain the fit and the automated procedure allows a quantitative analysis  
of a large number of spectra, which in the era of efficient multi-object 
spectrograph facilities is a very important aspect. We will describe the method in the 
subsections below. 

\subsection{Brief description of the model grid\label{grid}}

The basis for the quantitative spectral analysis is the same comprehensive 
grid of line blanketed LTE model atmospheres and very detailed non-LTE line formation 
calculations used by K08. For a detailed description of the physics of the model
atmosphere and line formation calculations see \citet{przybilla2006}.

Our grid of models covers a range of $\displaystyle 8300 \le T_{eff} \le 15000$~K  
(with steps of 250~K and 500~K below and above 10$^4$~K, respectively) and 
$\displaystyle 0.75 \le log~g \le 2.60$ (with steps of 0.05~dex) in effective temperature and 
gravity with the following metallicities at each grid point, 
[Z] = $\log\left(\mathrm{Z/Z}\right)_{\odot}$: -1.30, -1.00, -0.85, -0.70, -0.60, -0.50,
 -0.40, -0.30, -0.15, 0.00, 0.15, 0.30. The quantity Z/Z$_{\odot}$ is the metallicity 
relative to the Sun. The solar abundances were taken from \citet{grevesse98},
except for oxygen, which is from \citet{asplund2004}. Based on trend in the literature, a 
microturbulence of 8 \kms~is adopted for the low gravity models gradually changing to 4 \kms~at 
higher gravity. For further details we refer to K08. 

This grid was primarily designed for the analysis of Ia and Ib luminosity
class objects. Several of the WLM stars selected for the present work were classified by
\citet{bresolin2006} as luminosity class II. To consider these objects, we enlarged 
the grid with two new
effective temperatures, 8100 and 7900\,K, and extended to higher gravities all the models
with $\displaystyle T_{eff}\le 10^4$~K ($log~g \le 2.75$~dex).  

\subsection{General methodology}

The goal of the analysis is to determine for each star a set of four parameters 
$\{T_{eff}$, logg, [Z] and E(B-V)$\}$, for which the corresponding synthetic model spectrum 
best matches the observations. We have three data sets available: 
1) normalized and rectified spectra, which will give line profile information 
for the Balmer and metal lines; 2) flux calibrated spectra providing information 
about the Balmer jump, and 3) broad band photometry which can be used for the 
construction of the observed long wavelength SED.

\begin{figure}
  \begin{center}
  \includegraphics[width=0.80\textwidth,angle=90]{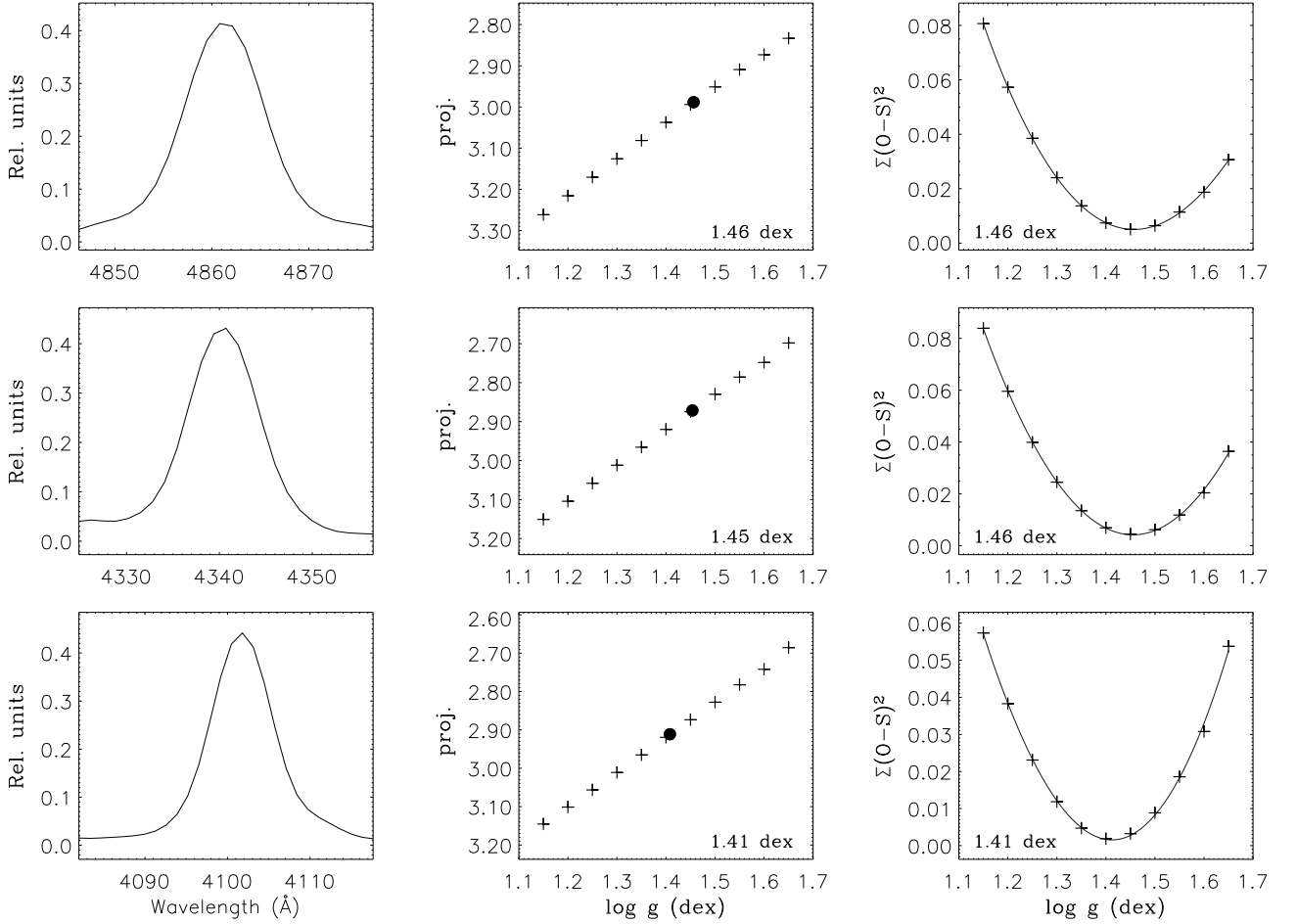}
  \caption{Determination of surface gravity. For illustration purposes, only three lines are
  presented. From top to bottom: H$\beta$, H$\gamma$ and
  H$\delta$. For each line, the first panel shows the base vector, the second one displays the projection of the
  models (crosses) and the observed spectrum (filled dot), and the third panel presents a solution by applying 
  a minimum distance method (minimization of the quadratic differences, 
  $\sum_{\lambda}\left(O_\lambda - S^j_\lambda\right)^2$ ) for comparison. The derived $log~g$ for each line is also
  included in each individual plot. \label{fig_proj_balmer}}
  \end{center}
\end{figure}

K08 have shown in detail how the theoretical spectral line profiles, the SED and 
the Balmer jump depend on temperature, gravity and metallicity. Generally speaking, 
the Balmer lines depend mostly on gravity, but show also a significant temperature 
dependence, whereas the Balmer jump depends mostly on temperature, but also on 
gravity. The dependence on metallicity of the Balmer lines and the Balmer jump is 
very weak. 
Using these properties, two different curves can be defined in the $T_{eff}$--$log~g$
plane, one being the locus of the models that reproduce the Balmer lines when the temperature 
is adopted, and, conversely, another tracing the locus of the models that reproduce the 
Balmer jump once the gravity is assumed. The intersection of these two curves  
yields the gravity and the temperature of the star observed. The classic method 
used by K08 is to construct these curves through a by-eye fit of the models to the 
observed data. In our new approach we have decided to use an automated numerical 
fit of the model spectra to the observed data.

a) Our algorithm is iterative and begins with the fit of the Balmer lines. We start 
with a first value of $T_{eff}$ and [Z] and try to find the model gravity $log~g$ 
which best matches the shape of the observed Balmer profiles. For this purpose, 
we carry out a Principal Component Analysis (PCA) of the normalized model spectra 
in spectral windows around the Balmer lines. For a fixed $T_{eff}$ and [Z] we 
have n models with different gravities at each grid point (n varies from 20 to 32 
depending on the grid point), each of them providing a set of normalized flux 
values S$_{j}(\lambda)$ (where the index j runs over the n models) around a selected 
Balmer line. These model flux values define a matrix $\mathbf{A}$ (with elements 
A$_{ij}\,=\,$S$_{j}(\lambda)$, where i labels the wavelength points and j runs over the 
models with different $log~g$). Following the concept of PCA, we determine the 
eigenvalues and the eigenvectors of the covariance matrix 
$\mathrm{cov}\!\left(\mathbf{A}\right)$ \citep{deeming1964,whitney1983} and 
identify the largest eigenvalue and its corresponding eigenvector as the 
principal component associated with the parameter $log~g$. Fig.~\ref{fig_proj_balmer} 
shows the typical eigenvectors $\mathbf{v}(\lambda)$ for three Balmer lines. 
As a consistency test we check that the cumulative percentage variance associated with this 
principal component is always above 95\% in all cases, meaning that 
more than 95\% of the information in the input data can be described by the principal 
component alone.

\begin{figure}
  \begin{center}
  \includegraphics[width=0.80\textwidth,angle=90]{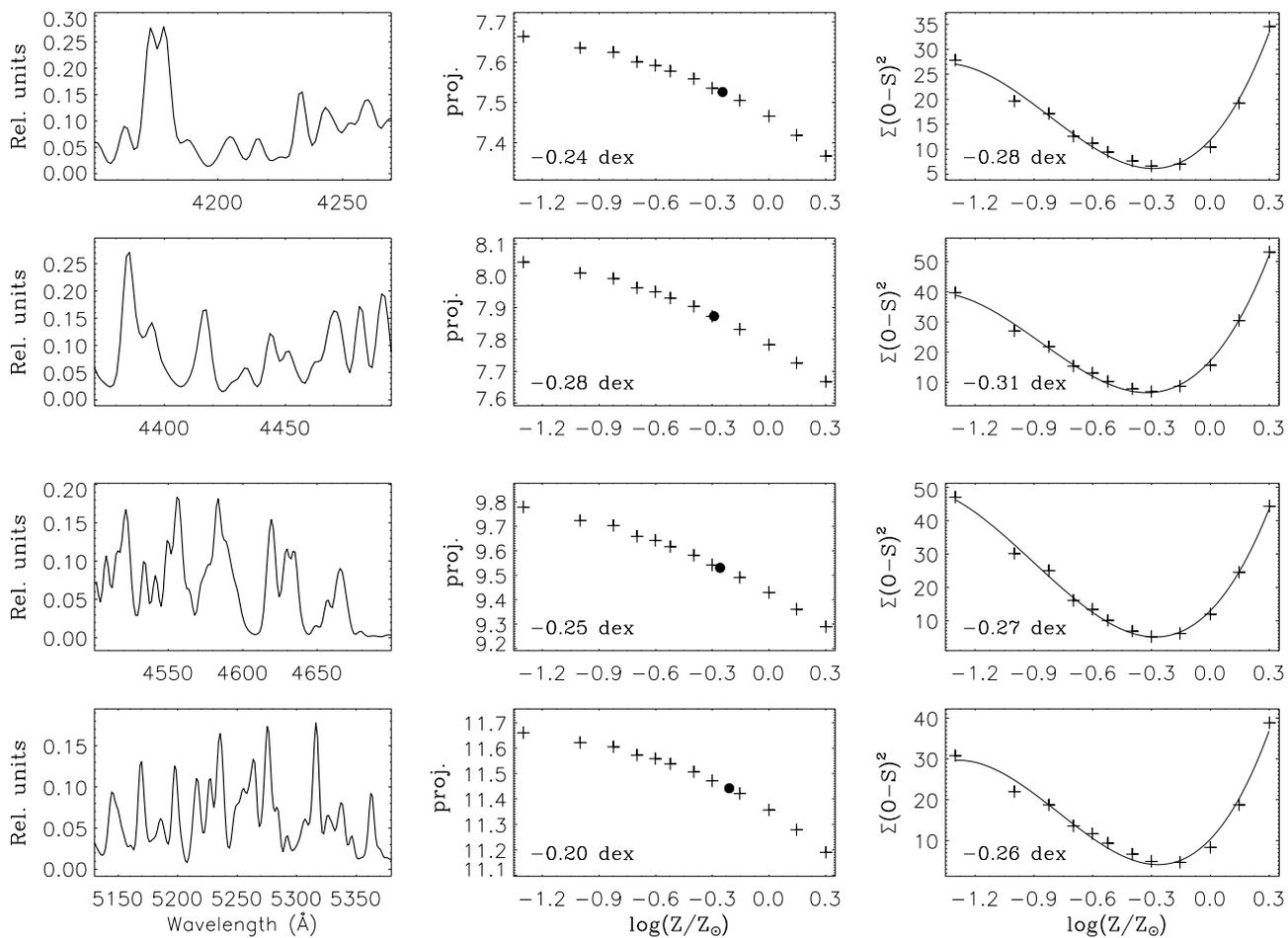}
  \caption{Metallicity determination. For illustration purposes, only four
  spectral windows are shown. For details about
  each individual panel, see previous figure. \label{fig_proj_met}}
  \end{center}
\end{figure}

The projection of the model flux matrix $\mathbf{A}$ onto the eigenvector $\mathbf{v}$ 
allows us to define a relationship between the $log~g$ values of the models
and their projections $\displaystyle\Phi(log~g)$, for each Balmer line
\[\displaystyle \sum_{\lambda}\mathrm{S}_{j}\left(\lambda\right)\,v\left(\lambda\right)\,=\,\Phi_{j}\,=\,\Phi\left(log~g\right) \]
which are also shown in Fig.~\ref{fig_proj_balmer}. Projecting the observed 
normalized flux values $F_\lambda$  in each Balmer line spectral window 
onto the corresponding eigenvector yields an observed value, which can then be 
compared to $\Phi(log~g)$ to find the best gravity $log~g$ at the selected values 
of $T_{eff}$ and [Z]. Comparing the values obtained from different Balmer lines 
(we usually use H$_{10}$ to H$_\beta$) allows us to assign a mean value and an 
uncertainty $\omega$ to this determination of the best fitting gravity.

b) Adopting this pair of $\left(T_{eff},\,log~g\right)$ we can now determine a new metallicity by defining 
spectral windows with many metal lines. Fig.~\ref{fig_proj_met} shows a typical example 
of different spectral windows and their corresponding PCA eigenvectors, this time determined 
with respect to metallicity [Z]. The projections $\Phi([Z])$ and the 
observed value are also shown. Using different spectral windows we can again assign 
an average value of [Z] and an uncertainty $\omega$. For this new [Z] we iterate the 
determination of $log~g$, but because of the weak dependence of the Balmer lines on 
metallicity changes are usually small.

c) In the previous steps we used Balmer and metal line profiles to obtain $log~g$ 
and [Z] (and their corresponding uncertainties) at a selected value of $T_{eff}$. Now 
we adopt these new values of $log~g$ 
and [Z] and use the observed SED around the Balmer jump to obtain a new value of $T_{eff}$. 
This is again done by PCA (but see below) as shown in Fig.~\ref{fig_proj_jump}. The model atmosphere 
SED corresponding to this $T_{eff}$, $log~g$ and [Z] is used to determine the reddening E(B-V) from a comparison 
($\chi^2$-minimization) with the observed broad band photometry, adopting the extinction law by 
\citet{cardelli1989}, as well as $\displaystyle R_v=A_V/E(B-V)=3.1$.
Since the region used to determine $T_{eff}$ could be affected by this extinction, we alternatively iterate
the values of $T_{eff}$ and E(B-V) until convergence in both is reached, defining the best $T_{eff}$ and the best
E(B-V). Unlike the cases of surface gravity and metallicity, the temperature is determined from
a single feature, the Balmer jump. The uncertainty in $T_{eff}$ is thus given by our ability to 
effectively measure the Balmer jump, and not from temperature values derived from different spectral 
windows. In 
order to estimate this uncertainty, we proceed in the following way. Two values of the temperature are
computed simultaneously, by using two different definitions of the Balmer
jump (see Fig.~\ref{fig_proj_jump}). The first one (Fig.~\ref{fig_proj_jump}d) corresponds to the 
value derived from the application of the PCA base vector (shown in Fig.~\ref{fig_proj_jump}c), while 
the second value (Fig.~\ref{fig_proj_jump}b) is based on the Balmer jump index $D_B$ introduced by K08, 
$\displaystyle D_{B}=\langle\log(F_{\lambda}^{long})\rangle-\langle\log(F_{\lambda}^{short})\rangle$, where  
\[\begin{array}{l}
\langle\log(F_{\lambda}^{short})\rangle=\lbrace\sum_{i=1}^{N}\log(F_{\lambda_{i}})\rbrace/N,~3585\bb\le\lambda_{i}\le3627\bb\quad\mathrm{,and}\\ 
\langle\log(F_{\lambda}^{long})\rangle=\lbrace\log F_{3782\bb}+\log F_{3814\bb}+\log F_{3847\bb}+\log F_{3876\bb}\rbrace/4  \nonumber
\end{array}\]
In the case of $\displaystyle F_{\lambda}^{short}$, $N$ is the number of wavelength points considered for the mean. Note 
that in both cases, we are considering all the models in the grid for a given pair $\displaystyle\left(log~g,[Z]\right)$, as 
previously explained. In an ideal case, with infinite S/N, both values would be 
exactly the same. Once noise is present, there is a difference between both temperatures, that reflects how 
precisely the Balmer jump can be measured, defining therefore the uncertainty in temperature, 
$\displaystyle\Delta T_{eff}$.

Apart from this $\Delta T_{eff}$, the difference between the initially assumed temperature
and the final derived value defines an error $\delta$, which can be assigned to 
the set of parameters $\displaystyle\{T_{eff}, log~g, [Z], E\left(B-V\right)\}$. This $\delta$ can 
be interpreted as the distance between the guess temperature and the true $T_{eff}$, because of the
weak dependence of the Balmer jump on gravity and metallicity.\\

We repeat the whole procedure (a, b, c) for different initial values of $T_{eff}$ and, 
thus, obtain a set of parameters 
$\pi_i\,=\,\{T_{eff}, log~g, [Z], \mathrm{E}\left(\mathrm{B-V}\right)\}_{i}$ 
together with their corresponding weights {$(\delta_{i}, \omega_{i})$}. The final solution  
$\Pi$ = $\{T_{eff}$, logg, [Z] and E(B-V)$\}$ is then obtained as a weighted mean  

\[ \Pi \,=\,\frac{\sum_{i}\,1/\delta^2_i\,\cdot\,1/\omega^2_i\,\cdot\,\pi_i}
{\sum_{j}\,1/\delta^2_j\,\cdot\,1/\omega^2_j} \]
where the indexes $i,j$ run over all the individual solutions for the corresponding parameter ($T_{eff}$, logg, 
[Z] or E(B-V)). In a similar way, a formal error (uncertainty) is obtained from 

\[ (\Delta\:\Pi)^2 \,=\,\frac{\sum_{i}\,1/\delta^2_i\,\cdot\,1/\omega^2_i\,\cdot\,(\pi_i - \Pi)^2}
{\sum_{j}\,1/\delta^2_j\,\cdot\,1/\omega^2_j} \]

This analysis method can only be applied in those cases for which the Balmer jump region 
is observed. In some instances, due to the location of the star in the observed field and its spectrum on the 
detector, this area is not observed, and
we have to rely on the normalized spectrum to obtain information about the effective temperature. In this situation
as already discussed by K08, the relative strength of lines from different species can be used. If the star
is hot enough to show \ion{He}{1} lines (earlier than around A0), these can be used to constrain the temperature. 
The accuracy is usually comparable to that when the Balmer jump is used. If 
the star is cool enough (later than B9), the relative behavior of different metal lines (for example \ion{Ti}{2} versus 
\ion{Fe}{2} lines, or the strength of the Fe-Mg blend located at $\sim\,$5150~\AA) could be used. To find the temperature, 
we then proceed in a similar way, defining several spectral windows in which these metal or helium lines are present, 
to construct the PCA eigenvectors (from the normalized synthetic spectra), but now with respect to temperature. There 
is an underlying assumption, though, when using this method: the adopted metallicity pattern. The implications of our adopted 
solar abundance pattern are discussed in Sect \ref{sect_micro}. Temperatures determined by this method have a
somewhat larger uncertainty, related to the accuracy with which the metallicity can be determined.

Metallicity and microturbulence are closely related with respect to the 
values derived from any analysis. As discussed by K08, it is difficult to constraint microturbulence  
at such low spectral resolution, since lines from a given species, with very different strengths, are 
required. In consequence, as explained in \S\ref{grid}, for each object, the microturbulence is given by
the final ($T_{eff}$,$log~g$) pair. We will discuss in \S \ref{sect_micro} the possible effects of the 
adopted relationship between the microturbulence and the ($T_{eff}$,$log~g$) values.

\subsubsection{Uncertainties}

The formal accuracy of the final solution depends primarily on the signal-to-noise ratio. Our experience shows
that, for the low spectral resolution considered in this work, a minimum S/N$\sim$50 is required for the metallicity 
determination in order to keep the uncertainties in a tolerable range (see next section), while T$_\mathrm{eff}$ and 
$log~g$ determinations, both based on stronger features (the Balmer jump and the Balmer lines) can still be carried 
out at much lower, S/N$\sim$15.

The formal errors arising from the application of our algorithm, as previously defined, are rather small. Typically, 
surface gravity and metallicity can be constrained to better than 0.05~dex, while the effective temperature 
uncertainty results in a few K. To define more meaningful errors, we also consider the 
uncertainties derived for the model in the grid with parameters closest  
to the final solution, i.e., we use the $\omega$ values (defined above) obtained for the set 
$\displaystyle\{T_{eff}, log~g, [Z]\}$ closest to the final solution. These provide an idea about how well the 
observations are reproduced by a single model.\\ 

The final errors quoted are obtained as a quadratic combination of these two sets of 
uncertainties. Typical values for these final errors result in $\sim$\,2--5\% in temperature, 
$\sim$\,0.1--0.2\,dex in surface gravity and $\sim$\,0.2--0.3\,dex in metallicity. As an example, 
Fig.~\ref{fig_classic_fit} shows the solution obtained for one of the stars in the sample, A14. 
The small box encloses the solution with the formal errors in the $T_{eff}$--logg plane. As can be seen 
this solution area agrees with the intersection of the conventional fit curves for the Balmer jump and
the Balmer lines, with the difference of the solutions well below the resolution limit of both methods. 
The final adopted errors are represented by the dashed box.

With regard to the uncertainties in the flux weighted gravity, it must be noted that the errors in $log~g$ and
$T_{eff}$ are correlated (see Fig.~\ref{fig_classic_fit}), which reduces the uncertainties in $log~g_\mathrm{F}$ in most cases. The reader is referred
to K08 for a detailed discussion on this topic.

\subsubsection{Goodness-of-fit assessment}

With the final set of parameters known for each star, it is useful to evaluate the fitness 
of each final solution, with the goal 
of identifying problems not detected in the analysis. Given a pair observation--final model 
$\displaystyle\left(O_\lambda,M_\lambda\right)$, we define the residuals of a spectral window
with $\displaystyle n_i$ wavelength points as  
\[\displaystyle r_i\left(M_\lambda,O_\lambda\right)=\frac{1}{n_i}\left[\sum_{j=1}^{n_p}\left(\frac{O_j-M_j}{M_j}\right)^2\right]^{1/2}\]
The fitness $\displaystyle R\left(\Pi\right)$ of a parameter $\Pi$ is given by the sum of these 
relative residuals $\displaystyle r_i$ over all the spectral windows 
considered in the determination of the corresponding parameter (as described above), weighted by 
the (normalized) S/N of each individual window 
\[R\left(\Pi\right)\:=\:\sum_{i=1}^{n_w}{r_i \hat{s_i} }\qquad \mathrm{ , with }\qquad \hat{s_i}=\frac{\mathrm{S/N}_i}{\sum_{k}{\mathrm{S/N}_k}} \]
In the case of the surface gravity and metallicity, 
the normalized spectra are used to calculate the fitness. For the effective temperature, the flux 
calibrated data and the synthetic SEDs are considered to evaluate the fitness when the Balmer jump 
is observed. If it is not observed the normalized spectra are used. 

Alternatively, we could define the fitness for each window as 
\[\displaystyle q_i\left(M_\lambda,O_\lambda\right)=\frac{1}{n_i}\sum_{j=1}^{n_p}\left(\frac{\log(O_j/M_j)}{\log(1+\epsilon)}\right)\]
where $\displaystyle\epsilon$ is the error allowed in the fitting (tolerance), defined as 
$\displaystyle\epsilon=\mathrm{S/N}^{-1}$. The global 
fitness, for each parameter individually, is given by the sum over all the windows, in this case without weighting by the S/N
since its effect is already accounted for with the tolerance. This second fitness evaluator has the
advantage of containing additional information in its sign, with a positive/negative value reflecting whether the 
corresponding parameter is over or under-estimated. This fitness definition is particularly useful to detect
variations in the goodness-of-fit from star to star. \\

We define the global fitness of the model with respect to the observation, 
$\displaystyle R\left(M_\lambda,O_\lambda\right)$, as the sum of the 
individual fitness values of each parameter, $\displaystyle R\left(\Pi\right)\!$. From the definitions, it is 
clear that the smaller 
the absolute value, the fitter the model. Fitness values corresponding to the final solutions are presented 
in Table~\ref{tab_fitness}, and they will be used in the following sections to discuss the results.

\subsection{Consistency tests}

We performed a number of simple tests to check the analysis algorithm. In all the following cases, we emulated the
observed spectra by degrading the models to a resolution of 5~\AA~(FWHM) and re-sampling them to 
1.32 \AA~(the spectral resolution and dispersion provided by FORS2 when equipped with the 600B grism). Note that 
these tests are meant to check our ability to reproduce known input parameters. 

First, we verified that, for any given model in the grid, we were able to recover almost exactly its parameters. In 
a second step, we created a number of models with parameters within the limits of the grid, but without any 
corresponding model in the grid. Three different sets of $\displaystyle T_{eff}$--logg,
representatives of a late B, an early A and a mid A supergiants were considered, for four different metallicities.
In all cases, we were able to recover the parameters to better than 1\% in effective temperature (based on the 
Balmer jump), 0.02~dex in surface gravity and 0.03~dex in metallicity. These two tests were 
carried out without considering noise, since they were intended to probe the ability 
of our algorithm to recover known parameters, without any influence from external sources.

A third set of tests was performed to check the effect of noise on the determination of metallicity, and hence to
evaluate the minimum S/N required for (meaningful) metallicity determinations. We selected a number of 
models in the
grid representative again of the different spectral groups, degraded them to different S/N (100, 50, 30 
and 15) and carried out a metallicity analysis. For each model and 
each S/N value, we did 100 independent trials. The results for the early A-type case are presented in Fig. 
\ref{fig_sigma_met} (the results are very similar for all the other $T_{eff}$--logg pairs considered). The first
two rows correspond to $\left[\mathrm{Z}\right]=0.00$~dex, while the other two rows are for 
$\left[\mathrm{Z}\right]=-0.85$~dex. Each single plot displays the relative frequency of occurrence 
versus the difference between the input and recovered metallicities, presenting also the sigma of the distribution. 
Since this sigma only describes the dispersion
introduced by the noise, the global uncertainty obtained in a generic analysis would be larger, once
the effects of the uncertainties in the other parameters (temperature and gravity) are accounted for. As a general
recipe, a maximum $\sigma_{S/N}\sim0.10$ dex at a fixed pair of $T_{eff}$ and $log g$ would be required in order to 
have a final global uncertainty $\sigma\sim0.20$ dex. 

Clearly, the minimum S/N depends on the metallicity: the higher the $\left[\mathrm{Z}\right]$, the stronger 
the lines, hence the lower the S/N required to detect them. At the same time, it also depends to some extent
on the spectral type; for a given metallicity, cool mid A-types (A1 to A3) present stronger
metal features 
(as well as a higher line density) than late B-/early A-types (B5 to A0). From Fig. \ref{fig_sigma_met}, it is
possible to identify a minimum S/N$\sim$50, in particular at low metallicities, to meet the requirement of a
global metallicity uncertainty around 0.2\,dex.

\subsection{Dependence of the results on hidden parameters\label{sect_micro}}

In order to keep the task of creating such a huge model grid manageable, only three parameters (the most
important ones, $T_{eff}$, $log g$ and [Z]) are explored so far. In the following we want to discuss some of the
possible implications regarding how the models are calculated.

As previously stated, metallicity and microturbulence are coupled in any analysis. Aimed at exploring 
the effect that our assumption about microturbulence $\xi$ (given for a pair $T_{eff}$-logg) 
could have on the results, we
calculated a small set of models with different values of $\xi$ (but the same $T_{eff}$, $log g$, [Z]  
considered for the grid), and analyzed them in the same manner as the observations. For a variation of 
$\xi$ of $\pm$2~\kms (which is the typical dispersion obtained in high resolution analyses of
these type of objects, for a given luminosity class) we found an error of less than 1\%, 0.01 dex and
0.07 dex for effective temperature, surface gravity and metallicity, respectively. As expected, both $T_{eff}$ 
and logg are unaffected by a change in microturbulence, and the larger effect is produced in the 
metallicity. Here we want to stress again that we are not dealing with individual lines, which certainly 
could present large variations with $\xi$ (like for example \ion{Mg}{2} 4481~\AA), but with all the spectral 
lines at once (or with as many of them as possible); the use of the whole spectrum weights the effect of the
turbulence differently, reducing its impact on the derived global metallicity. This is, on the other hand, a warning
sign for the use of small sections of the spectrum (at low resolution) to
determine individual elemental abundances, which would be seriously hampered by the unknown $\xi$.

We proceeded in a similar way in order to evaluate the impact of our assumption of a solar 
abundance pattern. We calculated a set of models for which the ratio of the $\alpha$-elements to Fe-group 
elements 
relative to the solar reference vary, with [$\alpha$/Fe] = $\pm\,$0.15, $\pm\,$0.30 dex (note that by 
definition, the value used in the grid is [$\alpha$/Fe] = 0.0 dex). The maximum change in effective 
temperature (based on the Balmer jump) and surface gravity due to these changes in relative abundances 
is less than 1\% and 0.01\,dex~respectively. The difference goes up to 6\% in temperature when using the
normalized spectrum instead of the Balmer jump. This result is not surprising since in this case, $T_{eff}$ is
based on the relative strength of different species, in particular of \ion{Ti}{2} to \ion{Fe}{2} lines, 
which depends on the relative abundances as well as on the temperature. As a byproduct, this exploratory 
study allowed us to refine the limits for some of the spectral windows considered in the determination 
of metallicities, allowing 
also to identify features that are produced mainly by Fe-group elements (Fe, Ni, Cr, \ldots) and, 
conversely, 
features produced by $\alpha$-elements (in particular Ti). This will be used in the future to evaluate the
possibility of extracting information from the spectra not only for the global metallicity but also for the
relative abundances of some prominent species.

\subsection{Comments on individual stars \label{sect_stars}}

In this section we present a brief discussion of the analysis of some of the targets. 

\begin{itemize}


\item A4: the Balmer jump region was not observed for this star, therefore we have to use the normalized spectrum
to estimate $T_{eff}$ from metal lines (the star is too cool to present He lines). In this situation, $T_{eff}$ is coupled to
[Z], the well know problem of the spectral type--metallicity degeneracy. Due to the combination of the low
metallicity and the relatively low S/N, the errors in [Z] and $T_{eff}$ are relatively large.

\item A5 and A17: the Balmer jump is missing for both stars. Fortunately, they are hot enough to present 
\ion{He}{1} lines, and can be used to constraint $T_{eff}$. Note that both stars have S/N above 50, which
greatly improves the situation with respect to the two previous cases.   

\item A6: the analysis of this star is problematic. While the determination of \Teff~and $log~g$ using the Balmer jump and the Balmer lines
is straightforward and results in fitness values similar as for the other stars, the determination of [Z] is not. The metallicity is not
well defined, as is indicated by the high values of $\displaystyle R\left([Z]\right)$ and 
$\displaystyle Q\left([Z]\right)$ compared to the
other stars. Moreover, the high value of [Z]=-0.15~dex is puzzling. We note that the spectral type found by 
\citet{bresolin2006} based on the relative strength of \ion{Ca}{2} H and K lines is A7, suggesting a significantly cooler temperature than the one obtained from the Balmer
jump. While there are other examples of A supergiants where the \ion{Ca}{2} related spectral types do not match \Teff~(see discussion of
star A16) mostly because of the contribution by interstellar lines (note that E(B-V) of A6 is 0.29~mag), we have tried to obtain a spectral
fit at cooler temperatures ($T_{eff}=8100$~K, $log~g=1.50$~dex) with a metallicity [Z]=-0.7~dex more in line with the other stars. At these 
parameters, the Balmer lines are reproduced well, but the fit of the Balmer jump is unacceptable as indicated by the fitness values in
Table~\ref{tab_fitness}. In addition, the fitnesses $R\left([Z]\right)$ and $Q\left([Z]\right)$ are only marginally improved and still
significantly worse than the fits for the other stars. The value for E(B-V) at this cool temperature is 0.27~mag, only slightly 
lower than before. 

The problem described above might be caused by stellar multiplicity. We have tried to emulate the observables (Balmer jump, normalized spectrum and broad band photometry) 
by combining two models, one hot to reproduce the Balmer jump and one cool to produce the rich metal spectrum and the Balmer lines. 
However, it has been impossible to find a combination of models to fit everything. Also, the star does not show signs of companions 
in archival WFPC2/HST images. Even more puzzling is the fact that all the different photometric measurements considered, from B- to 
K-band (including WFPC2/HST photometry in filters F555W and F814W taken from \citealt{holtzman2006}) are satisfactorily reproduced 
with a single model solution, invoking high reddening.

Another possibility is that we are dealing with a completely different type of star with just the photometric and (partial) spectroscopic
appearance of an A supergiant. The only solution that comes to our mind is a low mass star, for instance in post-RGB or post-AGB phase, in
the Galaxy. \citet{venn2003} have already discussed this possibility and found that this is rather unlikely. Indeed, if we assume the
typical mass of 0.5~M$_\odot$ for these objects, then with $log~g=1.95$~dex and $T_{eff}=8750$~K, we obtain $log~L/L_\odot=2.9$ or
$M_\mathrm{bol}=-2.5$~mag, with which, an apparent magnitude of $m_\mathrm{bol}=18.9$~mag would put the object at a distance of 190 Kpc,
certainly outside the Galaxy. We also note that A6 has a radial velocity typical for a WLM member \citep{bresolin2006}. 

At this point, we consider the metallicity of A6 obtained in our analysis as unreliable. We will include the object in our discussion of the
\fglr and of interstellar reddening, but will regard it uncertain. 


%
%
%
%
%


\item A14: this star has been previously studied by \citet{venn2003}, to which these authors referred
as WLM-15. In their analysis, \citet{venn2003} found a significant discrepancy between the oxygen
abundance of the star and the abundances obtained from \ion{H}{2} regions. We compare our results with those of 
\citet{venn2003}
in the following section. Unfortunately, our wavelength coverage does not include O lines, therefore it is 
not possible to carry out an independent analysis of the oxygen abundance.

\item A16: the spectral type classification by \citet{bresolin2006}, based on the \ion{Ca}{2} lines, is not consistent 
with the spectrum, nor with the results of the analysis (see below). There is at least another 
well known case for which this also happens. The Galactic star HD\,12953, the standard A1\,Ia in the MK system, would be
classified as an A3 based on the strength of its \ion{Ca}{2} lines, as already discussed by \citet{evans2003}. 
Unlike the case of A6, a completely consistent solution can be found. 


\end{itemize}
 
\section{Comparison with high resolution spectroscopy \label{sec_hires}}

One of the stars in our sample, A14, was previously studied by \citet[][hereafter V03]{venn2003}. These authors 
obtained a high 
resolution optical spectrum with UVES at the VLT (R$\sim$32000), and subsequently analyzed it using model 
atmosphere techniques similar to the ones employed here. This offers the possibility of a comparison with our low resolution work.

For the fundamental parameters, $T_{eff}$ and logg, V03 obtained 8300$\,\pm\,$200 K and 1.60$\,\pm\,$0.10 dex, while our solution is
8270$\,\pm\,$145 K and 1.60$\,\pm\,$0.12 dex. The agreement is very good. Worth noticing is the fact that the effective temperature of
V03 is based on the Mg ionization equilibrium, while ours is based on the Balmer jump. 
With respect to the chemical abundances, the high resolution work is far superior in the sense that it can provide 
individual elemental abundances,
while we derive a global metallicity. In order to compare with our result, we used V03's individual abundances of Fe, Ti, Sc and
Cr (and our solar references) and calculated a mean [Z] value of -0.44$\,\pm\,$0.13~dex (the 
latter number is the standard deviation of this weighted mean). Note that neither O nor 
N lines available to V03 are covered in our observed spectral range, thus we decided not to include them to compute 
[Z]. Within the uncertainties, this value is in good agreement with our derived value of -0.50$\,\pm\,$0.19~dex. In 
fact, this good concordance can be seen in Fig. \ref{fig_highres}, were we 
show the UVES spectrum of A14 and a model with the parameters resulting from our low resolution analysis,  
convolved with the appropriate UVES instrumental profile and rotational velocity.  

While the comparison can only be done for one star, the close agreement found is certainly very encouraging, in particular
given the relatively low metallicity of the object. This agreement is also 
in consonance with the result presented by \citet{bresolin2006} for the early B-type supergiant A9 in the same galaxy. 

\section{Results \label{sec_results}}

The stellar parameters obtained in our work are summarized in Table~\ref{tab_parameters}. In the following we discuss metallicity,
interstellar extinction and stellar parameters. 

\subsection{Metallicities}

The metallicity of the 6 BA stars in our sample range (excluding A6) from $\left[\mathrm{Z}\right]=-0.5$ 
to $-1.0$~dex. The weighted mean metallicity, is $\left[\mathrm{Z}\right]=-0.87\,\pm\,0.06$~dex, where the 
uncertainty is given by the standard deviation of the sample. This value compares well with the results of 
\citet{bresolin2006}, based on the $\alpha$-element content of three early B-type supergiants, with weighted 
mean of 
$\left[\mathrm{O/H}\right]=-0.86\,\pm\,0.07$ dex. It also agrees with the oxygen abundances obtained from \ion{H}{2}~regions 
\citep[][]{lee2005}. This 
agreement indicates that WLM's young stellar population while metal poor exhibits an abundance pattern very similar
to the Sun (at least with respect to the most relevant species, like O, Fe or Ti). 

It seems that A14's metallicity is an outlier, being 0.26\,dex above the mean, which is 
a bit larger than the individual uncertainties we are claiming in our analyses. 
As discussed above, our low resolution result is consistent with the high resolution analysis by 
\citet{venn2003}. The difference with respect to the rest of our sample,
along with the abnormally high oxygen abundance derived by \citet{venn2003}
could indicate that the star has a peculiar history. We note that K08 in their study of NGC\,300 have 
found two similar
outliers not representing the expected metallicity at their galactocentric distance. They expressed that 
the idea of homogeneous
metallicity might be naive. Excluding A14 from the weighted mean results in only a slight change to 
$\left[\mathrm{Z}\right]=-0.89\,\pm\,0.07$~dex.

Combining the results of our sample and the early B-type supergiants from \citet{bresolin2006}, 
there is no evidence of a metallicity dependence of the young stellar population with the 
spatial location in the galaxy (see Fig. \ref{fig_ebv_halpha}).

\subsection{Extinction}

Reddening values were determined using Johnson V-band and Cousins I-band photometry from
\citet{bresolin2006} \citep[see also][]{pietrzynski2007}, along with B-V and V-R colors from 
\citet{massey2007}. V- and I-band data 
for the stars in common in these papers are shown  
in Fig.~\ref{fig_photo}. For our sample of 11 B- and A-type supergiant stars (represented by filled 
dots in the figure) we obtain a difference in the zero points of $-0.03\pm0.02$~mag~and 
$-0.013\pm0.030$~mag~in V and I respectively, with the magnitudes from the first reference 
being fainter in both filters. Note that the star A17 was 
not considered in the I-band mean since it presents a large difference of almost -0.5~mag between
both datasets. Using multi-epoch photometry (around 100 epochs, spanning over 2 yr), \citet{bresolin2006}
did not detect variability, beyond the observational scatter, for any of our targets. It is thus unlikely that
the discrepancy
in A17 I-band magnitudes is related to variability. In the case of the early B-type supergiant A9,
its B-band magnitude is not consistent with all the other photometric measurements (see the corresponding 
plot in Fig.~\ref{fig_seds_ir}). In both cases, we did not take these values into account for the determination 
of the corresponding extinctions for the stars. 

For four of the stars in the sample (A4, A6, A9, and A10), we also have IR J- and K-band data from 
\citet{gieren2008}. 

The individual reddening values are presented in  Table~\ref{tab_parameters}, and some examples 
comparing the reddened synthetic SEDs with the photometry are shown in Fig.~\ref{fig_seds_ir}. 
Individual reddening values range from 0.03 to 0.12 mag, with the extreme case of 0.29 mag
for A6. Excluding this object, the mean reddening of the BA sample is 
0.07$\,\pm\,$0.01 mag, or 0.08$\,\pm\,$0.01~mag when considering also the 
three early B-type supergiants. This mean value is higher than the
characteristic foreground value of E(B-V)=0.037 mag \citep{schlegel1998}, indicating that 
our stars suffer from internal (WLM produced) reddening. Our mean value is in agreement with the 
recent result by \citet{gieren2008} based on multi-wavelength observations of 
Cepheids. These authors find a mean characteristic reddening of $0.082\pm0.02$~mag. 

With regard to A6, we note that a very high reddening value is derived for both the high 
temperature and the low temperature solution (discussed in Sect.~\ref{sect_stars}). This means that the star  
suffers three times the mean reddening. A6 could be spatially associated with, or close
enough to, one of the high column density \ion{H}{1} areas identified by \citet[][]{kepley2007}, in 
particular the region identified by these authors as the handle of the hook, extending north of 
the C1 H$\alpha$ complex of \citet{hodge1995}. However, two of the three early B-type 
supergiants, A9 and A10, seem to be also spatially related to high column density 
\ion{H}{1} regions, but they do not present such extreme reddening. We cannot discard, however, the possibility
of a patchy ISM on small scales, and comparable cases of high reddening have also been found by K08 in NGC\,300.

\subsection{Masses, radii and luminosities}

Once the fundamental parameters $T_{eff}$ and $log~g$ are known, it is possible to derive the distance dependent quantities 
by adopting
a distance to WLM. In the next section, we will derive a distance to the galaxy based on the Flux weighted 
Gravity--Luminosity Relationship, FGLR, of blue supergiant stars \citep{kudritzki2003}. For the purposes of this section, 
we adopt that distance, $\left(m-M\right)_0=24.99\pm0.10$~mag, without any further explanation, delaying 
its derivation to the next section. 

Using this distance, the apparent bolometric magnitudes can be converted to absolute bolometric magnitudes, and luminosities. 
From the luminosities and the effective temperatures, we compute the radii. With surface gravities and radii,
it is then possible to derive (spectroscopic) masses. All these quantities are summarized in Table~\ref{tab_masses}. To estimate
their uncertainties, we propagate the uncertainties in the fundamental parameters.

From the luminosities, we can also derive stellar masses using theoretical evolutionary models (see Fig.~\ref{fig_dhr}). For 
this purpose, we use
the mass-luminosity relationships presented by K08. In particular, we selected K08's fits to SMC metallicity models 
from \citet{maeder2001} and \citet{meynet2005} that take into account rotation. These evolutionary masses are 
given in the last column of Table~\ref{tab_masses}. The uncertainties of the spectroscopic masses take into account uncertainties 
in luminosities, gravities and radii. The uncertainties in evolutionary masses account only for the uncertainties in the derived 
luminosities and are much smaller. They are, however, affected by the systematic uncertainties of evolutionary tracks.
Fig.~\ref{fig_masses}a shows the comparison of evolutionary and spectroscopic masses, including the objects in NGC\,300 studied
by K08. Given the typical spectroscopic mass errors of $\pm0.1$ to $0.2$~dex, we conclude that we have agreement between the 
two types of mass determinations. However, there is an indication that the spectroscopic masses are systematically smaller than
evolutionary masses for high luminosities (above $M_\mathrm{bol}\sim-8$~mag, see Fig.~\ref{fig_masses}b). We 
will discuss this in section \ref{fglr_teo}.

From the direct inspection of the HRD and the masses given in Table~\ref{tab_masses}, we conclude that the BA stars of our sample were born
with masses in the range of 8 to 20 M$_\odot$, with the early B supergiants evolved from slightly more massive progenitors between $\sim$25
to 50 M$_\odot$. This is in agreement with the results found by K08 for NGC\,300, who explain the difference of the masses as a selection
effect. 
 

\section{Flux weighted gravity--luminosity relationship and distance to WLM \label{sec_fglr}} 

\citet[][see also K08]{kudritzki2003} revealed the existence of a tight correlation between the flux weighted gravity, 
$g_\mathrm{F}=g/T_\mathrm{eff}^4$, and the luminosity of BA supergiant stars. This relationship is supported by the predictions 
of evolutionary models. Very briefly, the physical reason behind this is that during their evolution from the Main Sequence,
massive stars with masses below $\sim60\,\mathrm{M}_\odot$ will evolve at almost constant luminosity and, due to the short 
timescale and the low mass loss rates, constant mass. In this case, the luminosity of the star is correlated with the 
flux weighted gravity. 

K08 have found a relation between the absolute bolometric magnitude and the flux weighted gravity of 
the form  
\begin{equation}
\label{eq1}
M_\mathrm{bol}=\mathrm{a}\left(\log g_\mathrm{F} - 1.5\right) + \mathrm{b}
\end{equation}
with their present best values of the coefficients a and b derived from a large sample of supergiants combining spectroscopic
results obtained for 8 different galaxies. These coefficients are given in Table~\ref{tab_fglr}. We also include the values  
obtained by K08 for only the supergiants in NGC\,300, to give an idea of the uncertainties involved in calibrating this relationship. 

The supergiants in WLM show a tight {\sc fglr} when plotted in apparent bolometric magnitudes 
(in Fig.~\ref{fig_fglr_mbol}). This can now be used to determine a distance. Following
K08, we fit an expression of the form \[m_\mathrm{bol}=\mathrm{a}\left(\log g_\mathrm{F} - 1.5\right) + \mathrm{b} \]
This fit is shown in Fig~\ref{fig_fglr_mbol} by the solid line, and the values for a and b are given in Table~\ref{tab_fglr} as well. 
With only ten stars the slope is somewhat uncertain, thus, we prefer to fix the slope to the value obtained by K08 for their large 
sample. We then recalculate the fit with this fixed slope (dashed line; the new value for the coefficient b is also presented 
in Table~\ref{tab_fglr}).  
The difference in b with respect to K08 then yields a first determination of the distance modulus, $\mu=\left(m-M\right)_0$, 
for which we obtain $\mu=24.99$~mag.

If we fix the slope to K08's \fglr for NGC\,300 stars only, and compare again, we derive $\mu=25.06$~mag. This can
be interpreted as an estimate of the possible systematic uncertainties to be about 0.06~mag.


The statistical uncertainty of the distance modulus is given by 
\[\sigma^2=\frac{1}{n\left(n-1\right)}\sum_{i=1}^{n}\frac{\left(M_{\mathrm{bol},i}-M^\mathrm{FGLR}_{\mathrm{bol},i}\right)^2}{\widehat{\sigma}_i^2} \] 
where $\displaystyle n$ is the number of WLM stars, $\displaystyle M^\mathrm{FGLR}_{\mathrm{bol},i}$ is obtained from 
Eq.~\ref{eq1} evaluated at 
the corresponding observed $\displaystyle g_{\mathrm{F},i}$,  
$\displaystyle M_{\mathrm{bol},i}$ is the apparent bolometric magnitude m$_{bol,i}$ corrected by the derived distance 
modulus $\mu$, and $\widehat{\sigma}_i=\sigma_i/\sum_{j}\sigma_j$, with 
$\displaystyle\sigma_i^2=\sigma_{m_\mathrm{bol}}^2 + a^2\sigma_{\log g_\mathrm{F}}^2$ the individual
uncertainties in m$_{bol}$, accounting for the uncertainty in the BC, in the extinction and in the observed 
photometric errors, as well as the uncertainties in the flux weighted gravity. We obtain a statistical uncertainty associated 
with the distance modulus of $0.10$~mag.

An alternative way to determine the distance modulus is to minimize the residuals in magnitudes once the stars are shifted to 
a particular distance, i.e. we determine the value of $\mu$ that minimizes 
\[S^2\!\left(\mu\right)=\frac{1}{n\left(n-1\right)}\sum_{i=1}^{n}\frac{\left(m_{\mathrm{bol},i}-\mu-M^\mathrm{FGLR}_{\mathrm{bol},i}\right)^2}{\widehat{\sigma}_i^2} \] 
For this second method, the uncertainty in the distance determination is given by the 
square root of the minimized residuals.

The distances derived by both methods are in excellent agreement. This is a consequence of the fact that the slope of the 
relationship defined by the WLM stars is very close to the slopes found by K08 (see Table\ref{tab_fglr}). Fig.~\ref{fig_fglr_final} 
presents the final distance corrected \fglr for the WLM stars (filled circles), together with all the objects 
used by K08 to calibrate the relationship. This
figure also includes K08's \fglr fit used to calculate the distance to WLM. 
As can be seen in this figure, the agreement is very good. 



Finally, we would like to point out that the solution used for A6 (cool or hot model) has little effect on the derived 
distance, since the values of the flux weighted gravity are very similar in both cases, and in any case, a change in the flux 
weighted gravity is correspondingly accompanied by changes in the bolometric correction and extinction (relatively minor in this case), in 
such a way that the star moves along the relationship. The physical reason behind this behavior of $g_\mathrm{F}$ has been 
extensively discussed by K08, to which the reader is referred for further insights.

\subsection{WLM FGLR: empirical versus theoretical relationships and metallicity dependence\label{fglr_teo}} 

In Fig.~\ref{fig_teofglr} we show the observed \fglr for WLM compared with the prediction of stellar evolution. We have chosen the stellar
evolution {\sc fglr}s for solar and SMC metallicity using evolutionary tracks by \citet{meynet2003}, and \citet{maeder2001} and 
\citet{meynet2005}, which include the effects of stellar rotation with initial rotational velocities of 300~\kms (for the related 
mass-luminosity relationships and the parameterization of the evolutionary {\sc fglr}s see K08).

We note that for low luminosity and high flux-weighted gravity ($log~g_\mathrm{F} > 1.6,\,M_\mathrm{bol} < -8$~mag) there is 
good agreement. However, towards lower gravities and higher
luminosities the stellar evolution {\sc fglr}s show a strong curvature and start to differ from the result of the spectral analysis. The
effect is stronger for SMC metallicity ([Z]=-0.7), which is close to the average metallicity we have determined for the young
stellar population in WLM. An indication of a similar discrepancy has already been noted by K08 (see their Fig.~27 and the corresponding
discussion). 

This discrepancy seems to be equivalent to a discrepancy between stellar evolutionary and spectroscopic mass at high luminosities (see
Fig.~\ref{fig_masses}). Note that (see K08, Eq.(29))
\[\displaystyle g_\mathrm{F}=g T_\mathrm{eff,4}^{-4}\propto M R^{-2} T_\mathrm{eff}^{-4}\propto M L^{-1}\]
and, in consequence, smaller spectroscopic masses result in a \fglr shifted towards smaller flux weighted gravities. Thus, one reason
for the discrepancy could be a systematic underestimate of stellar gravities by the spectral analysis at the high luminosity end of the
{\sc fglr}. We realize that the most discrepant objects in Fig.~\ref{fig_masses} and Fig.~\ref{fig_teofglr} are early B-type
supergiants, which where analyzed with a different stellar atmosphere code. On the other hand, K08 had a much larger sample of
supergiants available and did not find a systematic difference between high luminosity early B-types and BA-types.

Another possibility might be the effects of rotation on the evolutionary tracks. K08 noticed that tracks without rotation result in 
{\sc fglr}s shifted towards higher gravities because the mass-to-light ratio is larger. Thus, tracks with even higher initial rotation
than considered here may result in {\sc fglr}s in better agreement with the observations. However, such increase in 
rotational velocity will not be supported by recent studies of LMC and SMC Main Sequence B stars, progenitors of BA 
supergiants, showing projected rotational velocities below 
200 \kms~\citep[see for example][]{hunter2008}. 
\\

Finally, it is also important to discuss a possible metallicity dependence of the {\sc fglr}. Except for the very high luminosity
end, the relative difference between the evolutionary {\sc fglr}s for [Z]=0.0 and [Z]=-0.7 is very small. This difference is mostly 
caused by the effects of mass-loss, which are stronger at higher metallicity and higher luminosity. The fact that we obtain a distance 
consistent with the TRGB and Cepheid studies of WLM using the \fglr calibration by K08, which is based on objects mostly with LMC 
metallicity, supports the conclusion that metallicity effects are not important. Future work on supergiants in other metal poor 
galaxies will very likely help to clarify the situation.

\section{Conclusions}\label{sec_discussion}

The primary goal of this work has been to determine the distance to WLM using a new spectroscopic method, the {\sc fglr}. The distance
modulus obtained, 24.99$\pm$0.10~mag, compares well with recent determinations, based on purely photometric methods. \citet{rizzi2007}
found 24.93$\pm$0.04~mag from the TRGB \citep[see also][and references in Sect.~\ref{intro}]{pietrzynski2007} and the multi-wavelength study of Cepheids by
\citet{gieren2008} yielded 24.924$\pm$0.04$\pm$0.04~mag (statistical and systematic errors). With only 10 stars available for our 
study, the statistical uncertainty is larger than the ones claimed by the photometric studies. However, this could certainly be improved
with a larger sample of objects. Our study also confirms that the young population of WLM suffers from significant extinction, higher
than the foreground value. This is another example where the accurate measurement of intrinsic extinction turns out to be important for
the determination of distances. It is the advantage of the spectroscopic method presented here that it yields 
reddening and extinction for the individual targets. 

We have also determined stellar metallicities and found an average value of  [Z]=-0.87$\,\pm\,$0.06~dex, in good agreement with previous
\ion{H}{2} region studies. The fact that the TRGB and {\sc fglr} distances agree at this low metallicity indicates that metallicity
effects are very likely small for the \fglr~method.

Stellar evolution calculations can be used to determine a theoretical {\sc fglr}, which can then be compared with 
observations. We find
good agreement at low luminosities independent of metallicity assumptions for the theoretical relationship, but 
disagreement at higher 
luminosities with the theoretical {\sc fglr}, which are more important for the low metallicity. Assuming higher 
initial rotational velocities, which would enhance mass-loss and rotational mixing, might be a way to solve 
this discrepancy. These higher rotational rates are not supported by recent studies of the progenitors of these objects
in the Magellanic Clouds.

Finally, our results for A14 (aka WLM-15) are in good agreement with the results obtained by \citet{venn2003}, 
confirming to some 
extent the particular nature of this object. Unfortunately, we could not perform an independent analysis of its oxygen abundance. 

In summary, this first application of the {\sc fglr} method to determine a distance to a galaxy has proven to be successful. The {\sc
fglr} technique seems to be a robust and reliable way to provide independent and accurate information about extragalactic distances.

\acknowledgments
WG and GP gratefully acknowledge financial support for this work from the Chilean Center for Astrophysics FONDAP 15010003, and
from BASAL Centro de Astrof\'{\i}sica y Tecnolog\'{\i}as Afines (CATA). Support from the Polish grant N203 002 31/046 and the
FOCUS subsidy of the Foundation for Polish Science (FNP) is also acknowledged. We would like to warmly
thank Joachim Puls for his careful reading of the manuscript and suggestions. Finally, the anonymous
referee is acknowledge for his constructive comments.

{\it Facilities:} \facility{VLT (FORS2)}.

\appendix
\section{Synthetic photometry} 
In this appendix we present specific details relative to our synthetic photometry, that we consider could
be useful to others.

We follow the ideas presented by \citet[][in particular see its Section 1.6]{bessell2005} in order to
compute the photometric magnitudes from our model atmosphere models. Given a spectral energy distribution
$f_\lambda$, the magnitude $\displaystyle m_A$ in a given filter with transmission curve 
$\displaystyle R_A\left(\lambda\right)$ is calculated accounting for
the fact that modern detectors count the number of photons, not energy, therefore
$\displaystyle m_A \propto \int f_\lambda \left(\lambda\, R_A\left(\lambda\right)\right) d\lambda $. 

In order to determine the zeropoints of the different bandpasses, we use the recent spectral energy
distribution of Vega presented by \citet{bohlin2007}, along with the filter curves obtained from different
references. Table~\ref{tab_synthephot} contains the zeropoints of our synthetic photometry, as well as the
references for the filter transmission curves.

\clearpage


\begin{deluxetable}{l c c l r c c c r r r }
 \tabletypesize{\scriptsize}
 \tablecaption{WLM stars analyzed in this work. Primary identification following \citet{bresolin2006}.
 Alternative identification and photometry from \citet{massey2007} is also included. For completeness, information
 about the three early B supergiants studied by \citet{bresolin2006} is presented in this table. All photometric quantities
 are in magnitude units. \label{tab_objects}}
 \tablewidth{0pt}
 \tablehead{ 
 \colhead{ID} & \colhead{V} & \colhead{V-I} & \colhead{Spectral Type} & S/N &  & 
 \colhead{Alt. ID} & \colhead{V} & \colhead{B-V} & \colhead{V-R} & \colhead{R-I} \\ 
 \cline{1-5} \cline{7-11} 
 \multicolumn{5}{c}{(Bresolin et al. 2006)} & & \multicolumn{5}{c}{(Massey et al. 2007)} }
 \startdata 
 
A6  &  19.82 &  0.38 & A7\,Ib   &  49 & &  0156.16-152624.5 & 19.789 &  0.217 &  0.179 &  0.203 \\ 
A4  &  20.22 &  0.07 & A2\,II   &  44 & &  0201.57-152527.0 & 20.185 &  0.016 &  0.038 &  0.056 \\ 
A14 &  18.43 &  0.23 & A2\,II   &  96 & &  0159.56-152926.1 & 18.374 &  0.087 &  0.073 &  0.120 \\ 
A16 &  18.44 &  0.16 & A2\,Ia   &  96 & &  0157.89-153013.1 & 18.383 &  0.204 &  0.058 &  0.055 \\ 
A2\tablenotemark{a}  &  20.16 &  0.09 & A0\,II   &  44 & &  0159.04-152442.8 & 20.142 & -0.014 &  0.052 &  0.027 \\  
A12 &  17.98 &  0.06 & B9\,Ia   & 119 & &  0153.22-152839.5 & 17.966 &  0.005 &  0.031 &  0.014 \\ 
A5  &  19.41 & -0.04 & B8\,Iab  &  64 & &  0203.31-152552.6 & 19.412 & -0.117 & -0.021 & -0.082 \\ 
A17 &  19.34 &  0.00 & B5\,Ib   &  67 & &  0200.81-153024.8 & 19.313 & -0.109 & -0.026 &  0.451 \\ 
A9\tablenotemark{b}  &  18.44 & -0.06 & B1.5\,Ia & 101 & &  0157.20-152718.0 & 18.392 &  0.201 & -0.028 & -0.059 \\ 
A10\tablenotemark{b} &  19.34 & -0.15 & B0\,Iab  &  68 & &  0154.06-152745.4 & 19.317 & -0.171 & -0.050 & -0.093 \\ 
A11\tablenotemark{b} &  18.40 & -0.18 & O9.7\,Ia & 106 & &  0159.95-152819.0 & 18.378 & -0.109 & -0.069 & -0.120 \\ 
\enddata
\tablenotetext{a}{This stars is outside the limits of the grid of models, therefore it cannot be analyzed.}
\tablenotetext{b}{Early B-type supergiant studied by \citet{bresolin2006}.}
\end{deluxetable}
\clearpage

\begin{deluxetable}{l c c c c }
 \tabletypesize{\scriptsize}
 \tablecaption{Mean J and K photometry from multi-epoch observations by \citet{gieren2008}\label{tab_IR}}
 \tablewidth{0pt}
 \tablehead{ 
 \colhead{ID} & \colhead{J} & \colhead{$\sigma\left(\mathrm{J}\right)$} & \colhead{K} & 
 \colhead{$\sigma\left(\mathrm{K}\right)$} \\
 \colhead{} & \colhead{(mag)} & \colhead{(mag)} & \colhead{(mag)} & \colhead{(mag)} 
 } 
 \startdata 
A6  & 19.055 & 0.022 & 18.896 & 0.129 \\
A4  & 19.889 & 0.021 & 19.843 & 0.067 \\
A2  & 19.916 & 0.010 & 19.748 & 0.158 \\  
A9  & 18.578 & 0.015 & 18.709 & 0.111 \\
A10 & 19.624 & 0.009 & 19.673 & 0.082 \\
\enddata
\end{deluxetable}
\clearpage



\begin{deluxetable}{lcccccrrr}
\tabletypesize{\scriptsize}
\tablewidth{0pt}
\tablecaption{Fitness of the solutions. See text for definitions of $\displaystyle R$ and $\displaystyle Q$. 
Stars are grouped in two sets, according to the method used to determine the temperature. For the first group, the Balmer jump is used,
and for the second we use the \ion{He}{1} lines for A5 and A17, and the relative strength of \ion{Ti}{2} lines for A4 (see text). 
\label{tab_fitness}}
\tablehead{
\colhead{ID} & \colhead{$\displaystyle R\left(log~g\right)$} & 
\colhead{$\displaystyle R\left([Z]\right)$} & \colhead{$\displaystyle R\left(T_{eff}\right)$} &
\colhead{$R$} & 
\colhead{S/N} & \colhead{$\displaystyle Q\left(log~g\right)$} &
\colhead{$\displaystyle Q\left([Z]\right)$} & \colhead{$\displaystyle Q\left(T_{eff}\right)$}
}
\startdata											  
A6\tablenotemark{a}& 0.1273 & 0.1397 & 0.0799 & 0.3469 &  49  & -0.3480 &  4.2911 & -0.9223 \\			   
A6\tablenotemark{b}& 0.1190 & 0.1435 & 0.0213 & 0.2838 &  49  &  1.0856 &  4.3364 & -0.0874 \\			
A14                & 0.1085 & 0.0430 & 0.0210 & 0.1726 &  96  & -0.6183 &  0.2116 & -0.0461 \\				 
A16                & 0.0651 & 0.0535 & 0.0037 & 0.1223 &  96  & -0.3119 &  0.3723 &  0.0067 \\
A12                & 0.0397 & 0.0388 & 0.0196 & 0.0982 &  119 & -1.0653 &  0.0388 &  0.0675 \\				 
\cline{1-9}   
A4                 & 0.1404 & 0.1047 & 0.0836 & 0.3288 &  44  &  0.2761 & -1.7217 & -0.0375 \\	       
A5                 & 0.0641 & 0.0768 & 0.0689 & 0.2099 &  64  & -0.7499 &  0.8870 &  0.0169 \\
A17                & 0.0842 & 0.0479 & 0.0305 & 0.1682 &  67  & -0.5869 &  0.2315 &  0.0004 \\					 
\enddata							     	 			  
\tablenotetext{a}{Cool solution, does not reproduce the Balmer jump, see text} 				  
\tablenotetext{b}{Hot solution that reproduces the Balmer jump} 				 
\end{deluxetable}										 
\clearpage

%
%

\begin{deluxetable}{crcccccc}
\tabletypesize{\scriptsize}
\tablewidth{0pt}
\tablecaption{WLM - Stellar parameters \label{tab_parameters}}
\tablehead{
\colhead{ID} & \colhead{$T_{eff}$}  & \colhead{$log~g$}  &
\colhead{$log~g_\mathrm{F}$} & \colhead{[Z]}  & \colhead{E(B-V)} & \colhead{$\xi$} & \colhead{comments}\\
\colhead{}     &  \colhead{(K)}         & \colhead{(cgs)}      &
\colhead{(cgs)}	      & \colhead{}     &  \colhead{(mag)} & \colhead{(\kmsb)} &\colhead{} } 
\startdata
A6   &  8750$\pm$240 & 1.95$\pm$0.17 & 2.19$\pm$0.12 & -0.15$\pm$0.33 & 0.295$\pm$0.010 &  4 & \tablenotemark{a} \\    
A4   &  8550$\pm$350 & 2.00$\pm$0.20 & 2.27$\pm$0.13 & -0.77$\pm$0.26 & 0.060$\pm$0.030 &  4 & \tablenotemark{b} \\     
A14  &  8270$\pm$144 & 1.60$\pm$0.12 & 1.93$\pm$0.09 & -0.50$\pm$0.19 & 0.124$\pm$0.007 &  4 & \\  		    
A16  & 10650$\pm$120 & 1.78$\pm$0.05 & 1.67$\pm$0.03 & -0.70$\pm$0.23 & 0.112$\pm$0.023 &  6 & \\  		    
A12  & 12100$\pm$136 & 1.79$\pm$0.03 & 1.46$\pm$0.01 & -0.78$\pm$0.21 & 0.061$\pm$0.004 &  8 & \\  		    
A5   & 12220$\pm$430 & 2.20$\pm$0.07 & 1.87$\pm$0.02 & -0.70$\pm$0.14 & 0.030$\pm$0.010 &  4 & \tablenotemark{c} \\    
A17  & 13500$\pm$450 & 2.35$\pm$0.07 & 1.86$\pm$0.03 & -1.00$\pm$0.15 & 0.061$\pm$0.004 &  5 & \tablenotemark{c} \\    
A9   & 20000$\pm$1000& 2.45$\pm$0.10 & 1.62$\pm$0.03 & -1.00$\pm$0.20 & 0.130$\pm$0.020 & 12 & \tablenotemark{d} \\ 
A10  & 25000$\pm$1000& 2.90$\pm$0.10 & 1.62$\pm$0.03 & -0.80$\pm$0.20 & 0.160$\pm$0.020 & 15 & \tablenotemark{d} \\  
A11  & 29000$\pm$1000& 3.00$\pm$0.10 & 1.62$\pm$0.03 & -0.80$\pm$0.20 & 0.070$\pm$0.019 & 15 & \tablenotemark{d} \\ 
\enddata
\tablenotetext{a}{Metallicity uncertain, see text}
\tablenotetext{b}{no Balmer jump measured; $T_{eff}$ determined from spectrum (Ti~{\sc ii} lines, see text)}
\tablenotetext{c}{no Balmer jump measured; $T_{eff}$ from HeI lines, see text}
\tablenotetext{d}{Results from \citet{bresolin2006}, with E(B-V) updated with the IR photometry as well as \citet{massey2007} data.}
\tablecomments{Note that the microturbulence is not derived in this work. See text for an explanation.} 

\end{deluxetable}
\clearpage

\begin{deluxetable}{crcccrrr}
\tabletypesize{\scriptsize}
\tablewidth{0pt}
\tablecaption{Bolometric corrections and distance dependent magnitudes: radii, luminosities and masses \label{tab_masses}}
\tablehead{
\colhead{name}    &
\colhead{$m_{V}$} &  \colhead{BC} & \colhead{$m_{bol}$} & \colhead{$\log L/L_\odot$} &
\colhead{R} & \colhead{$M^{\mathrm{spec}}$}     & \colhead{$M^{\mathrm{evol}}$}     \\
\colhead{}        & 
\colhead{(mag)}	  & \colhead{(mag)} & \colhead{(mag) \tablenotemark{a}}  & 
\colhead{(cgs) \tablenotemark{b}} &  
\colhead{(R$_{\odot}$)\,\tablenotemark{b}} &
\colhead{(R$_{\odot}$)\,\tablenotemark{b}} & 
\colhead{(M$_{\odot}$)\,\tablenotemark{b}}
} 
\startdata
A6  &  19.82$\pm$0.03 & -0.02$\pm$0.02 & 18.85$\pm$0.04 & 4.35$\pm$0.05  &  65.5$\pm$5.4 & 14.3$\pm$6.1 & 10.2$\pm$0.4 \\  
A4  &  20.22$\pm$0.05 & -0.05$\pm$0.05 & 19.98$\pm$0.12 & 3.90$\pm$0.09  &  40.7$\pm$5.3 &  6.1$\pm$3.2 &  7.7$\pm$0.4 \\  
A14 &  18.43$\pm$0.02 &  0.00$\pm$0.01 & 18.05$\pm$0.03 & 4.67$\pm$0.05  & 106.3$\pm$7.2 & 16.4$\pm$5.1 & 12.8$\pm$0.5 \\  
A16 &  18.44$\pm$0.02 & -0.43$\pm$0.01 & 17.65$\pm$0.07 & 4.83$\pm$0.07  &  76.8$\pm$6.4 & 13.0$\pm$2.6 & 14.4$\pm$0.8 \\  
A12 &  17.98$\pm$0.01 & -0.70$\pm$0.01 & 17.07$\pm$0.03 & 5.06$\pm$0.05  &  77.8$\pm$5.1 & 13.6$\pm$2.0 & 17.4$\pm$0.8 \\	      
A5  &  19.41$\pm$0.02 & -0.81$\pm$0.03 & 18.59$\pm$0.08 & 4.46$\pm$0.07  &  38.1$\pm$4.1 &  8.4$\pm$2.3 & 11.0$\pm$0.5 \\	   
A17 &  19.34$\pm$0.02 & -0.85$\pm$0.02 & 18.28$\pm$0.07 & 4.58$\pm$0.07  &  38.9$\pm$4.1 & 11.3$\pm$3.0 & 11.9$\pm$0.6 \\	   
A9  &  18.44$\pm$0.02 & -1.91$\pm$0.12 & 16.13$\pm$0.18 & 5.44$\pm$0.11  &  43.9$\pm$7.1 & 19.8$\pm$7.9  & 26.8$\pm$2.7 \\
A10 &  19.34$\pm$0.02 & -2.46$\pm$0.10 & 16.39$\pm$0.21 & 5.34$\pm$0.12  &  25.0$\pm$4.0 & 18.1$\pm$9.2  & 24.5$\pm$2.7 \\
A11 &  18.40$\pm$0.02 & -2.81$\pm$0.09 & 15.38$\pm$0.17 & 5.74$\pm$0.11  &  29.5$\pm$4.0 & 31.8$\pm$11.6 & 35.9$\pm$4.0 \\

\enddata
\tablenotetext{a}{Apparent bolometric magnitude:  $\displaystyle m_{bol}$ = ($\displaystyle m_{V}$ - $\displaystyle A_{V}$) + BC}
\tablenotetext{b}{Distance dependent magnitudes evaluated for $\displaystyle \left(m-M\right)_0\,=\,24.99\,\pm\,0.10$~mag}
\end{deluxetable}
\clearpage

\begin{deluxetable}{rccl}
\tabletypesize{\scriptsize}
\tablewidth{0pt}
\tablecaption{FGLR coefficients,  
$\displaystyle \{ M_\mathrm{bol},m_\mathrm{bol}\}\,=\,a\left(log~g_\mathrm{F} - 1.5\right) + b$. 
\label{tab_fglr}}
\tablehead{
\colhead{Relationship}  &  \colhead{a} & \colhead{b} & \colhead{Comments} } 
\startdata
K08-all      & 3.41$\pm$0.16 & -8.02$\pm$0.04  & $M_\mathrm{bol}$, calibrating \fglr using stars in 8 galaxies  \\
K08-NGC\,300 & 3.52$\pm$0.25 & -8.11$\pm$0.07  & $M_\mathrm{bol}$, calibrating \fglr using only NGC\,300 stars  \\
WLM          & 3.48$\pm$0.36 & 16.95$\pm$0.12  & $m_\mathrm{bol}$, WLM stars analyzed in this work             \\ 
WLM-all      & {\it 3.41}    & 16.97$\pm$0.09  & $m_\mathrm{bol}$, WLM stars with \fglr slope from K08-all      \\
WLM-NGC\,300 & {\it 3.52}    & 16.95$\pm$0.08  & $m_\mathrm{bol}$, WLM stars with \fglr slope from K08-NGC\,300 \\
\enddata
\tablecomments{Adopted slopes are shown in italics} 
\end{deluxetable}
\clearpage

\begin{deluxetable}{crrl}
\tabletypesize{\scriptsize}
\tablewidth{0pt}
\tablecaption{Synthetic photometry zeropoints \label{tab_synthephot}}
\tablehead{
\colhead{Bandpass}  &  \colhead{Zeropoint} & \colhead{$\lambda_0$} & \colhead{Reference} \\
\colhead{}          &  \colhead{ (mJy) }   & \colhead{ (\AA) }     & \colhead{}           }      
\startdata
V                   &   3647.62   &  5450 & \citet{bessell1990}  \\
I                   &   2432.91   &  7980 & \citet{bessell1990}  \\ 
B (KPNO)            &   4070.71   &  4381 & NOAO filter K1002 \tablenotemark{a} \\
V (KPNO)            &   3726.61   &  5387 & NOAO filter K1003 \tablenotemark{a} \\
R (KPNO)            &   3100.14   &  6513 & NOAO filter K1004 \tablenotemark{a} \\
I (KPNO)            &   2488.43   &  8245 & NOAO filter K1005 \tablenotemark{a} \\
J (2MASS)           &   1594.00   & 12350 & 2MASS web page \tablenotemark{b}  \\
Ks (2MASS)          &    667.00   & 21590 & 2MASS web page \tablenotemark{b}  \\
\enddata
\tablenotetext{a}{http://www.lsstmail.org/kpno/mosaic/filters/filters.html}
\tablenotetext{b}{http://spider.ipac.caltech.edu/staff/waw/2mass/opt\_cal/}
\end{deluxetable}
\clearpage 


\clearpage

\clearpage

\clearpage
\begin{figure}
  \begin{center}
  \includegraphics[scale=.7]{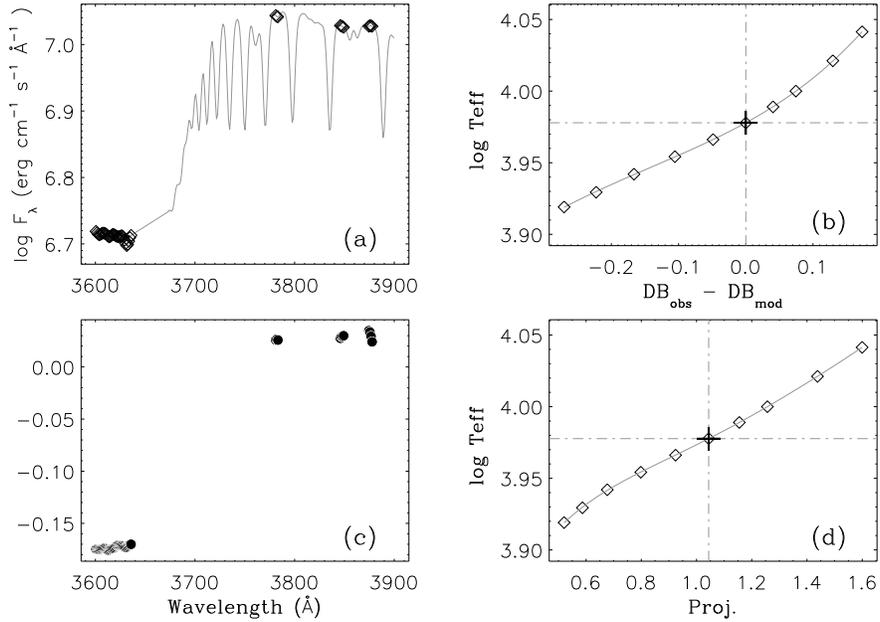}
  \caption{Determination of $T_{eff}$: (a) definition of the Balmer jump region; diamonds identified the wavelength points use
  to create the PCA base vector. (b) $T_{eff}$ determination based on the minimum difference between the observed and modeled Balmer 
  jump measurements, with 
  D$_\mathrm{B}\,=\,<\log$F$_{\lambda}^\mathrm{pre}>\,-\,<\log$F$_{\lambda}^\mathrm{post}>\,$ following K08; 
  the cross marks the derived temperature. (c) PCA base vector for the Balmer jump, and (d) projection of the models 
  (hollow diamonds) and the observed Balmer area (cross). \label{fig_proj_jump}}
  \end{center}
\end{figure}

\clearpage
\begin{figure}
  \begin{center}
  \includegraphics[]{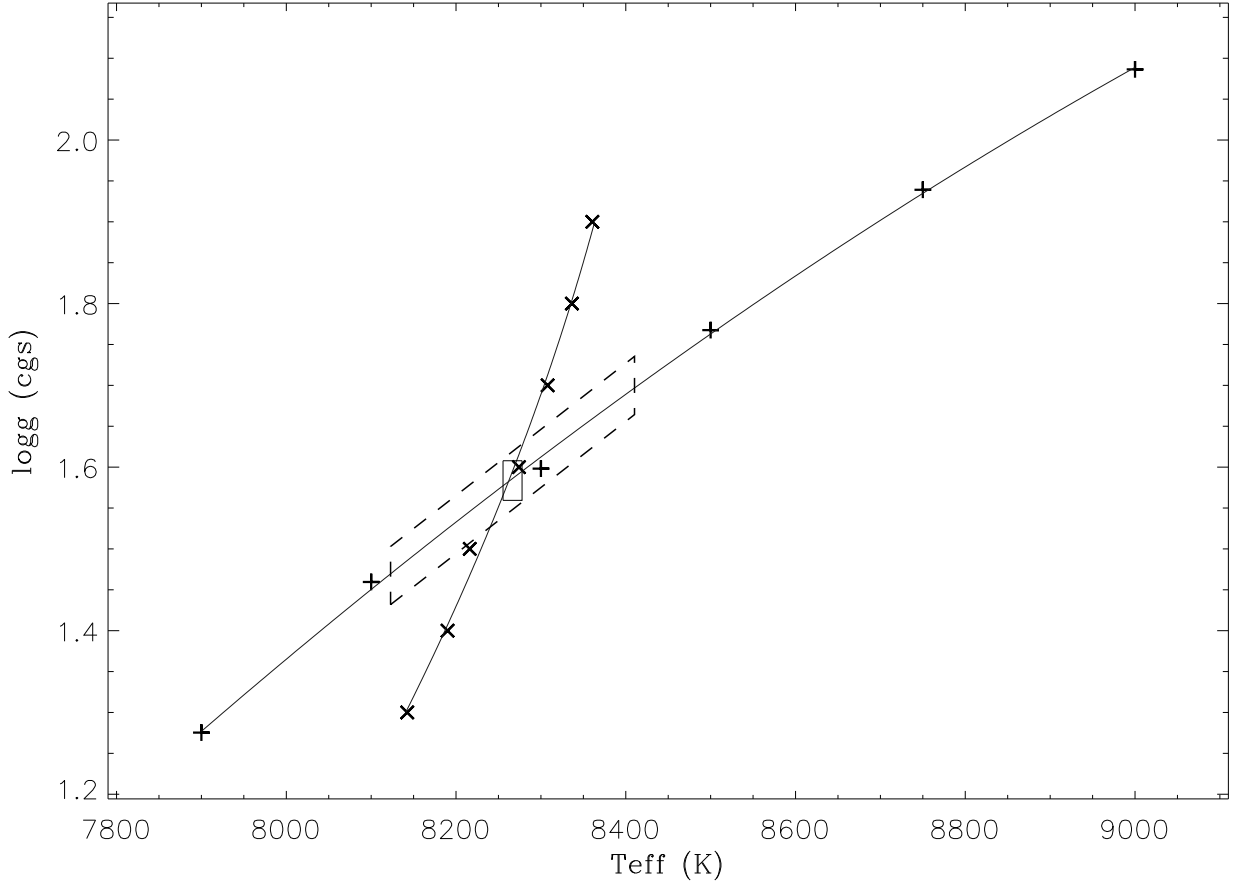}
  \caption{$T_{eff}$--$log~g$ diagram for star A14 of our sample. The solution provided by the algorithm 
  is defined by the small box, while the solution obtained by the conventional method would be located at 
  the intersection of the two $T_{eff}$--$log~g$ fit curves for the Balmer jump (x) and the Balmer lines (+). 
  The dashed lines define the final uncertainties in the derived parameters.     
  \label{fig_classic_fit}}
  \end{center}
\end{figure}

\clearpage
\begin{figure}
  \begin{center}
  \includegraphics[width=0.80\textwidth]{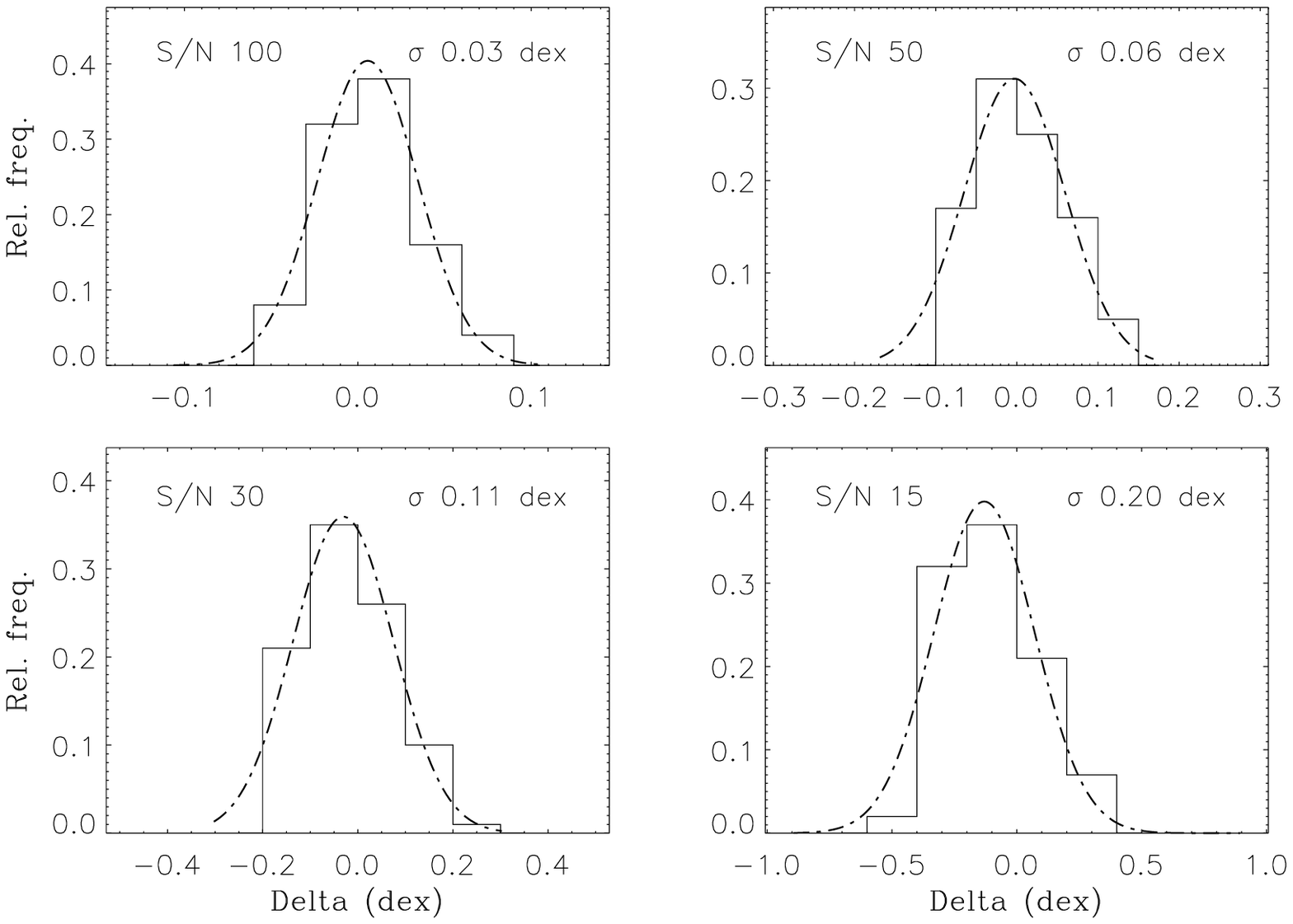}
  \includegraphics[width=0.80\textwidth]{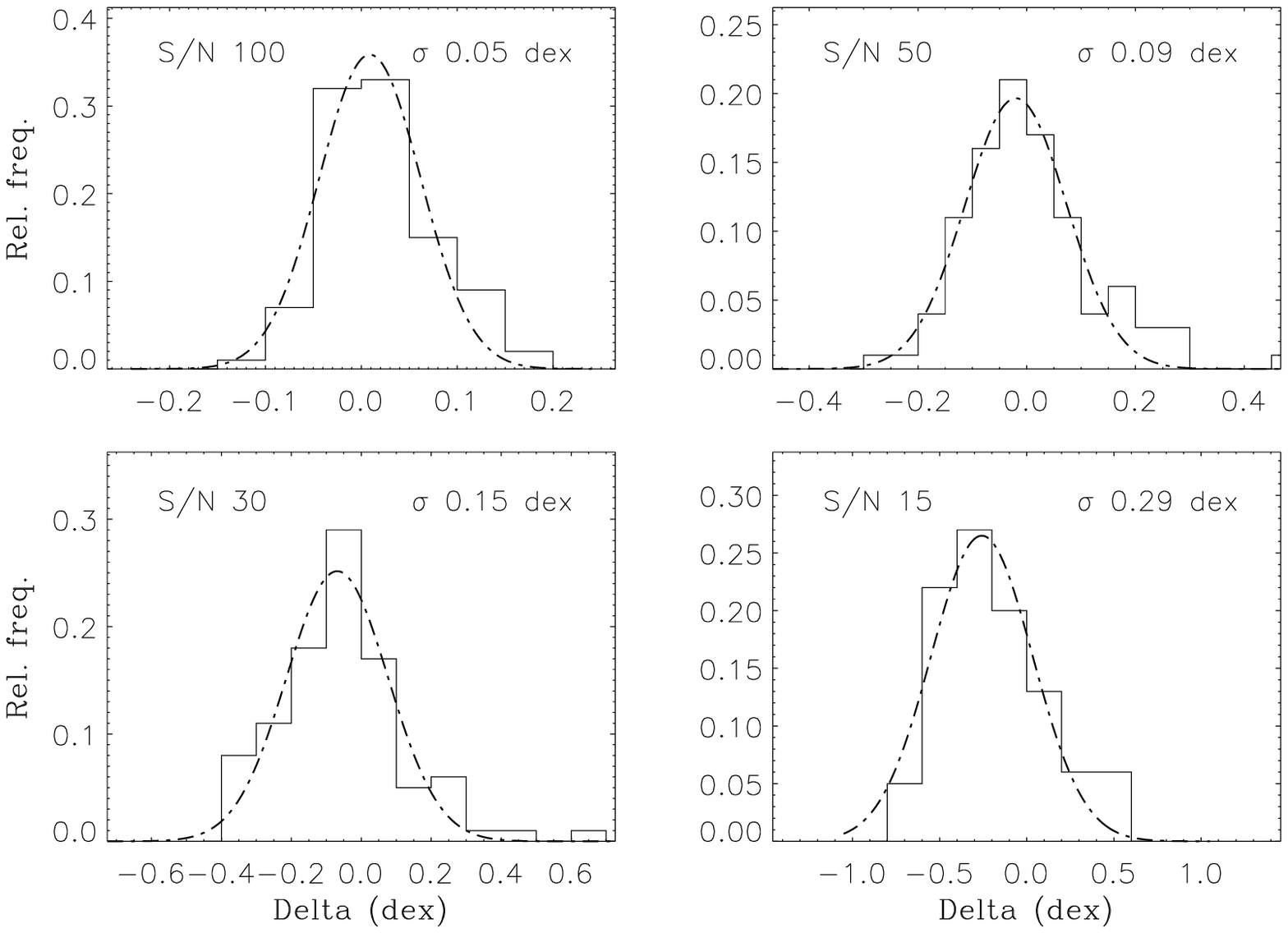}  
  \caption{Relevance of the S/N on metallicity determination. First four plots correspond to a 
  $\left[\mathrm{Z}\right]= 0.00$~dex case; the second set of plots present the case for 
  $\left[\mathrm{Z}\right]=-0.85$~dex. In each individual plot, the x-axis 
  represents the difference between input and derived metallicities, and the y-axis gives the relative frequency, for a 
  total number of 100 independent trials. A Gaussian fit to the resultant distribution is shown, and its sigma 
  is given in each plot. 
  \label{fig_sigma_met}}
  \end{center}
\end{figure}

\begin{figure}
  \begin{center}
  \includegraphics[width=0.90\textwidth,angle=180]{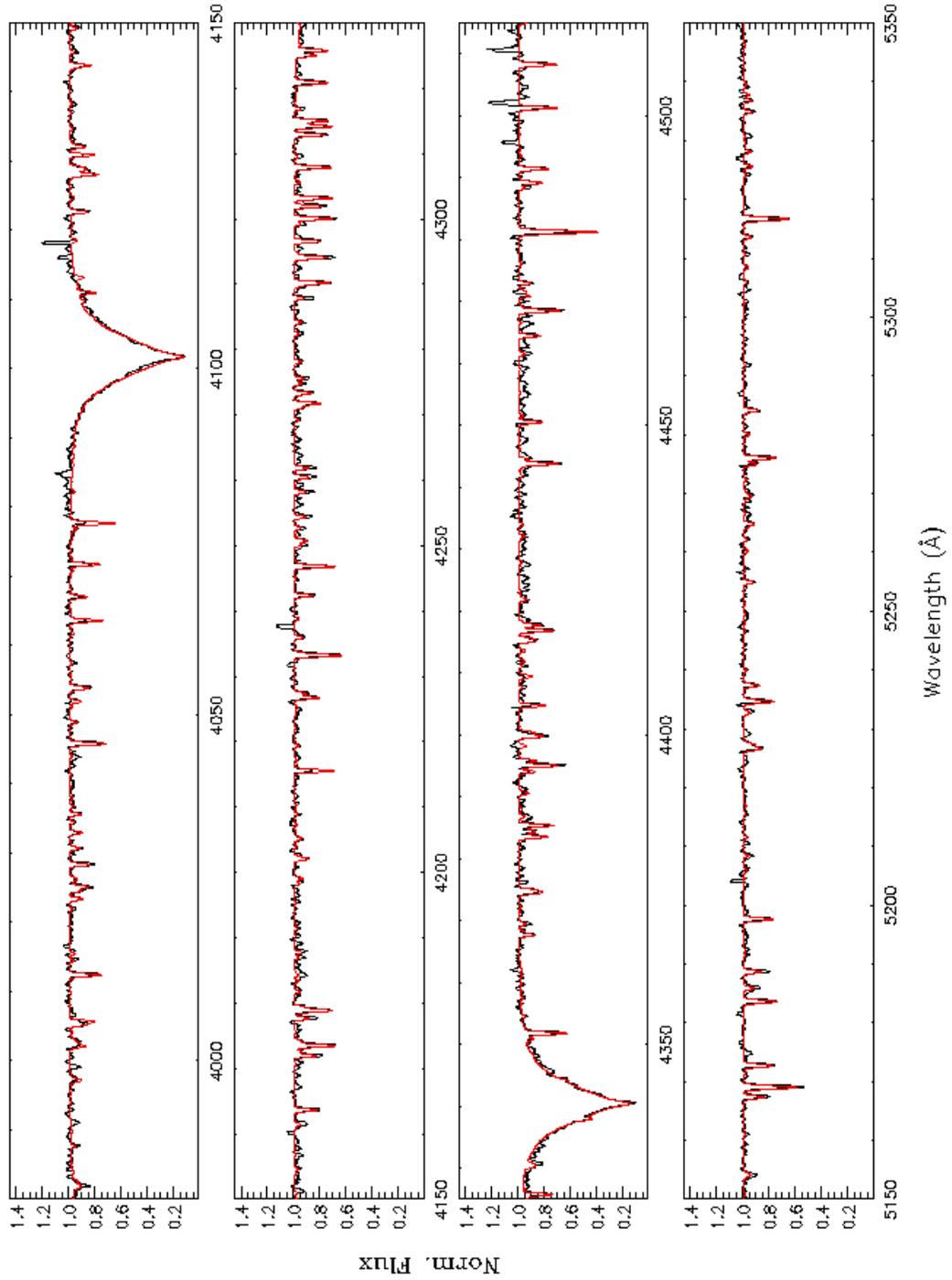}
  \caption{UVES/VLT (R=32000) spectrum of WLM-A14 compared with our solution (red) 
  obtained from the analysis of low-resolution FORS2/VLT (R$\sim$1000) spectrum. 
\label{fig_highres}}
  \end{center}
\end{figure}

\clearpage
\begin{figure}
  \begin{center}
  \includegraphics[width=0.90\textwidth]{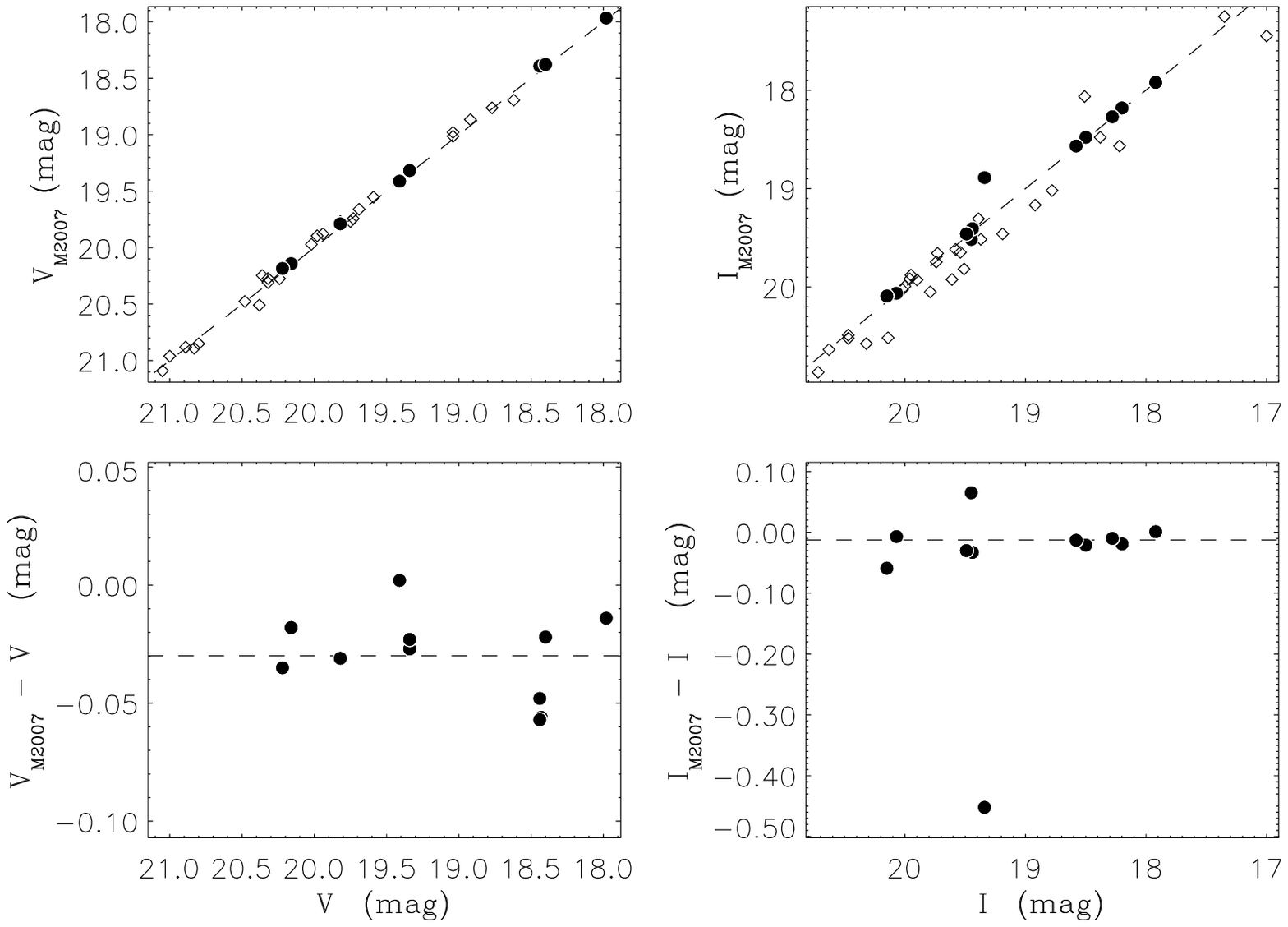}
  \caption{Comparison of photometric measurements. First row: V- and I-band photometric 
  data from \citet{bresolin2006} and \citet[][M2007 subindex]{massey2007}, for a total number of 35 stars in 
  common (diamonds). The stars analyzed in this work, and the three early B-type supergiants of 
  \citet{bresolin2006} are represented by filled circles. The
  1:1 relationship is represented by the dashed line. Second row: considering only the 11 stars analyzed 
  in our work, we find a mean difference in the zero point calibration of $-0.03\pm0.02$~mag~and 
  $-0.013\pm0.030$~mag~in V and I respectively. The star A17, not considered in the I-band mean,
  presents a difference with respect to our reference value of almost -0.5~mag.
  \label{fig_photo}}
  \end{center}
\end{figure}

\clearpage
\begin{figure}
  \begin{center}
  \includegraphics[width=0.90\textwidth]{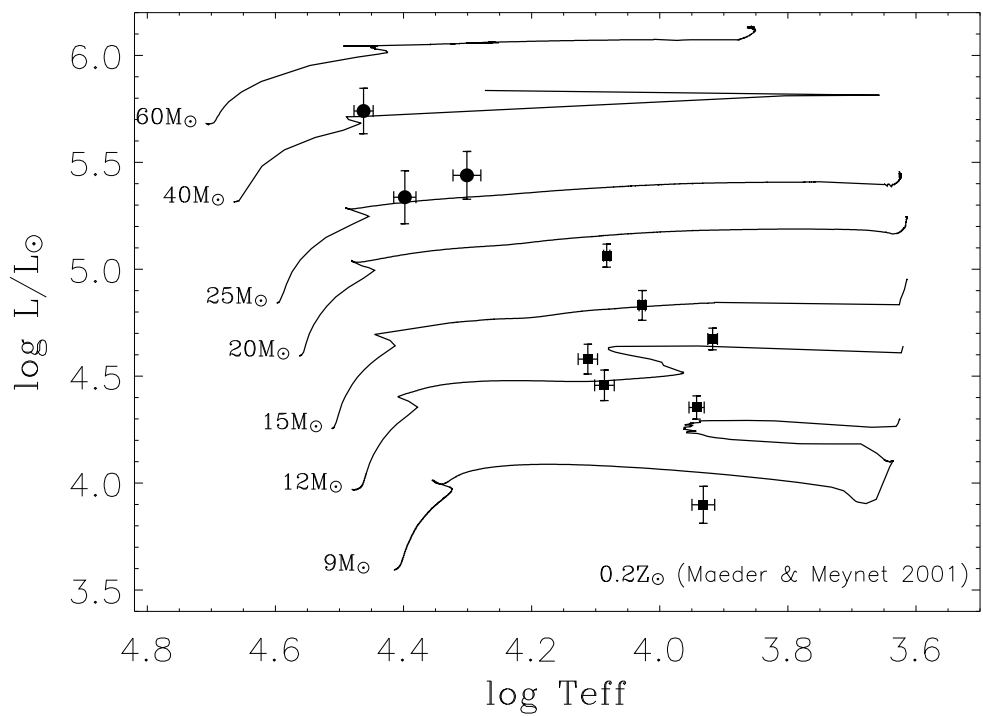}
  \caption{WLM Hertzsprung-Russel diagram. Circles locate the three early B-type supergiants analyzed by
  \cite{bresolin2006}, once corrected for the difference in distance modulus. The filled squares display the
  sample of BA supergiants analyzed in this work. Evolutionary tracks for rotating models from \citet{maeder2001} 
  are also shown, labeled with their initial mass. \label{fig_dhr}}
  \end{center}
\end{figure}

\clearpage
\begin{figure}
  \begin{center}
  \includegraphics[width=0.70\textwidth]{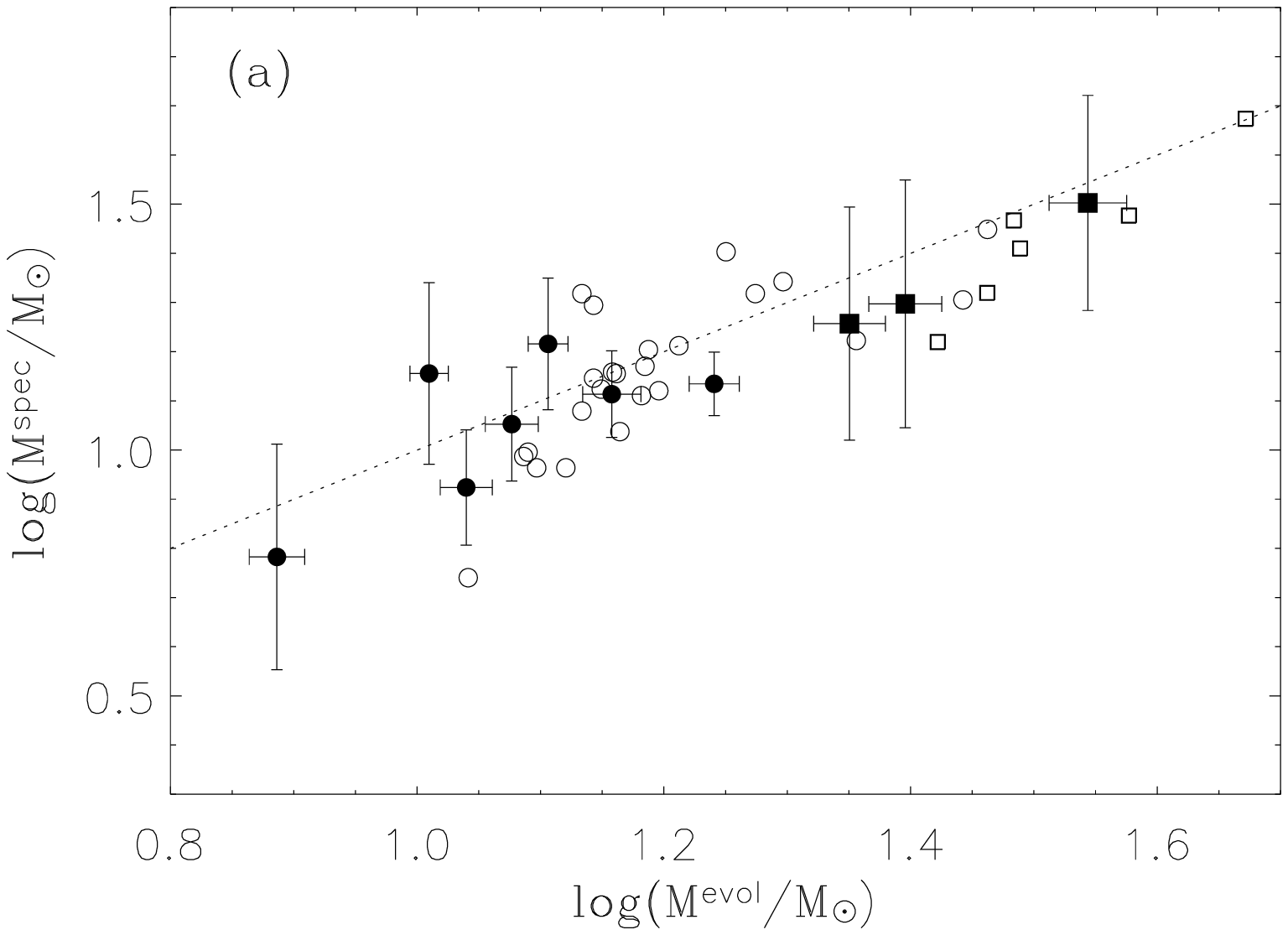}
  \includegraphics[width=0.70\textwidth]{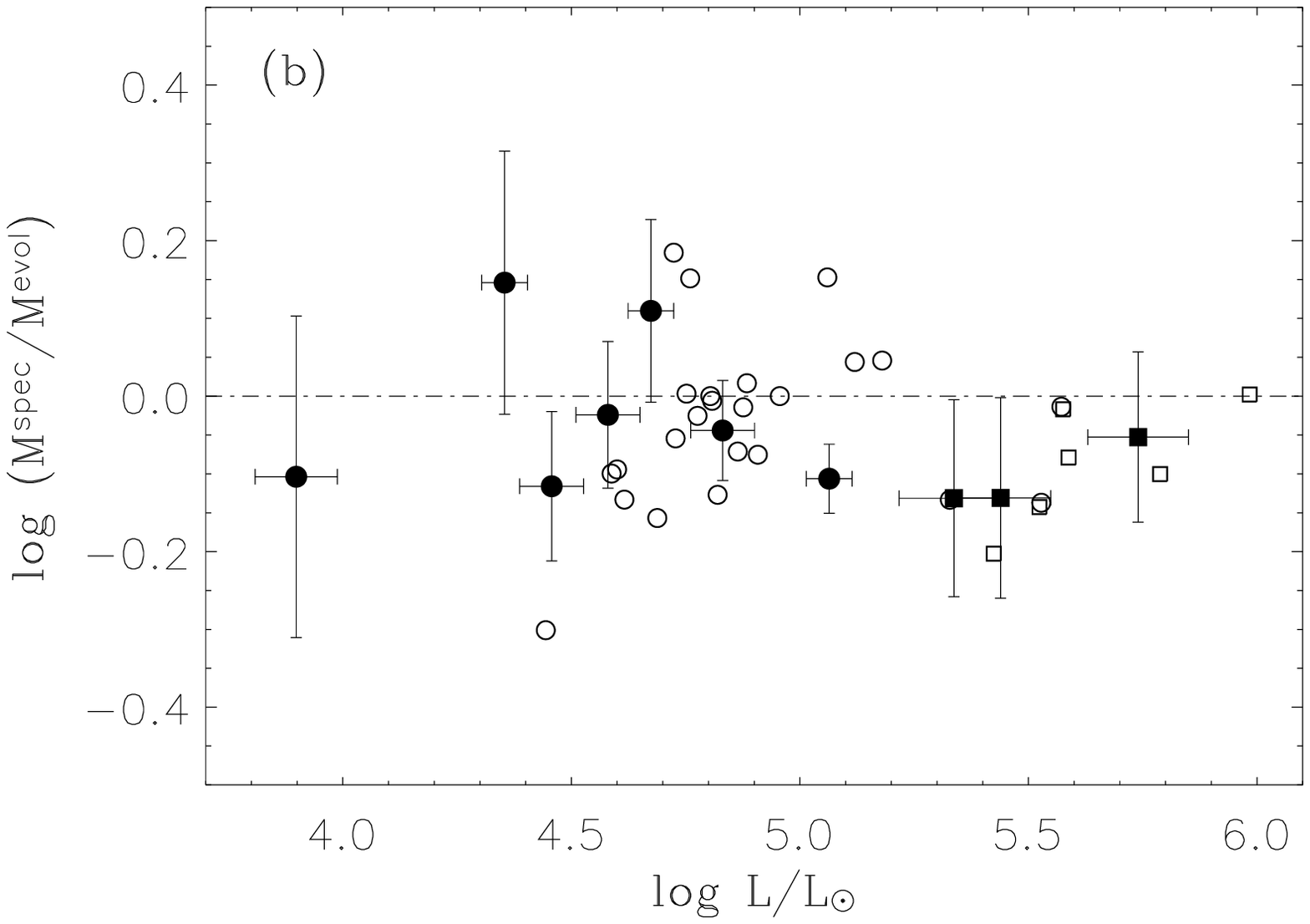}
  \caption{WLM stellar masses. (a) Comparison of evolutionary and spectroscopic masses for our sample of 
  WLM stars (filled circles) and the B supergiants of \citet[][filled squares]{bresolin2006}. The sample of NGC\,300 BA supergiants 
  from K08 (circles) as well as the early B supergiants from \citet[][empty squares]{urbaneja2005} are also shown, with the 
  dotted line defining the 1:1 relation. (b) The ratio of spectroscopic to evolutionary mass versus the stellar luminosity, for the WLM sample.
  Symbols as in (a).\label{fig_masses}}
  \end{center}
\end{figure}

\clearpage
\begin{figure}
  \begin{center}
  \includegraphics[scale=.45]{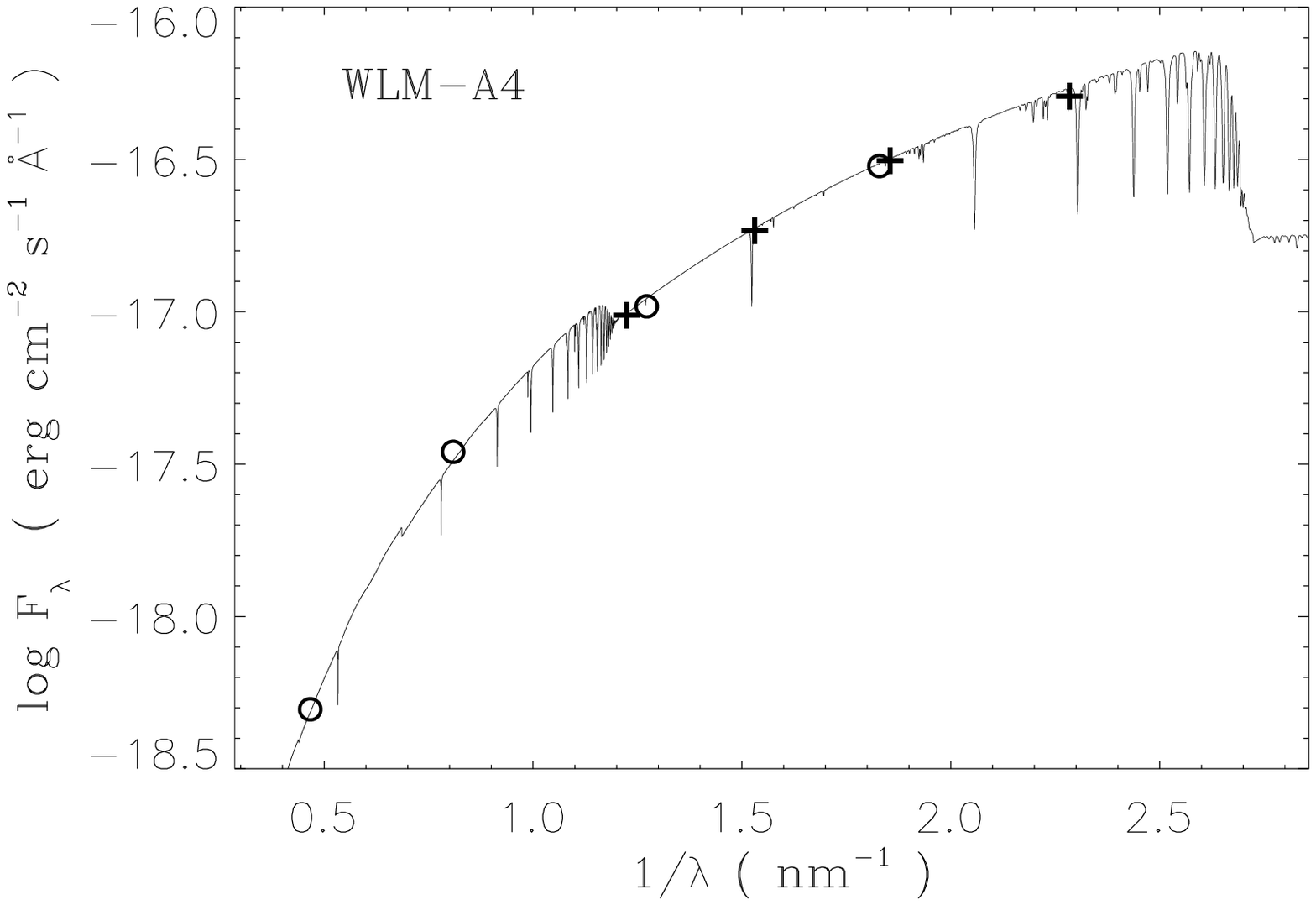}
  \includegraphics[scale=.45]{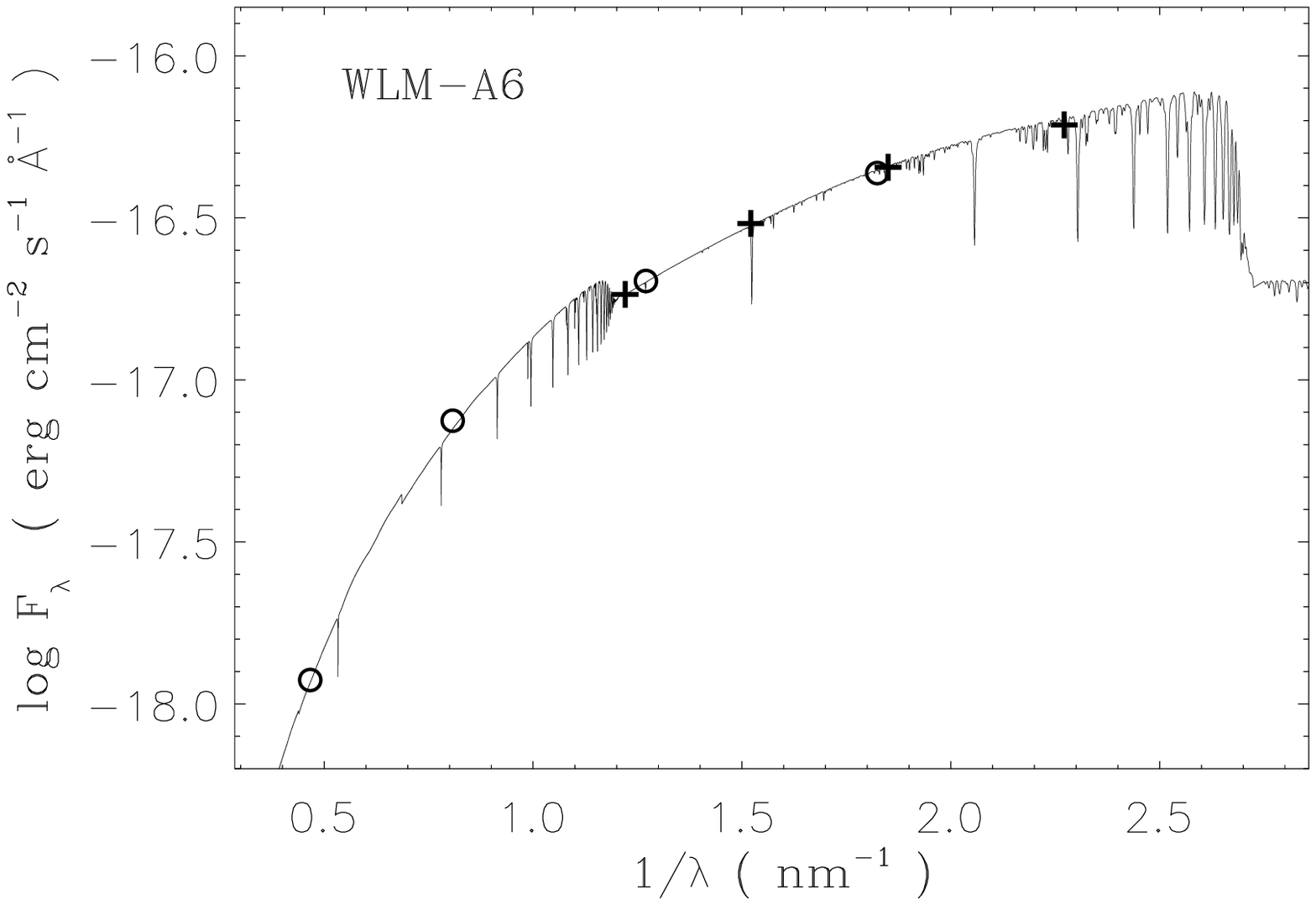}
  \includegraphics[scale=.45]{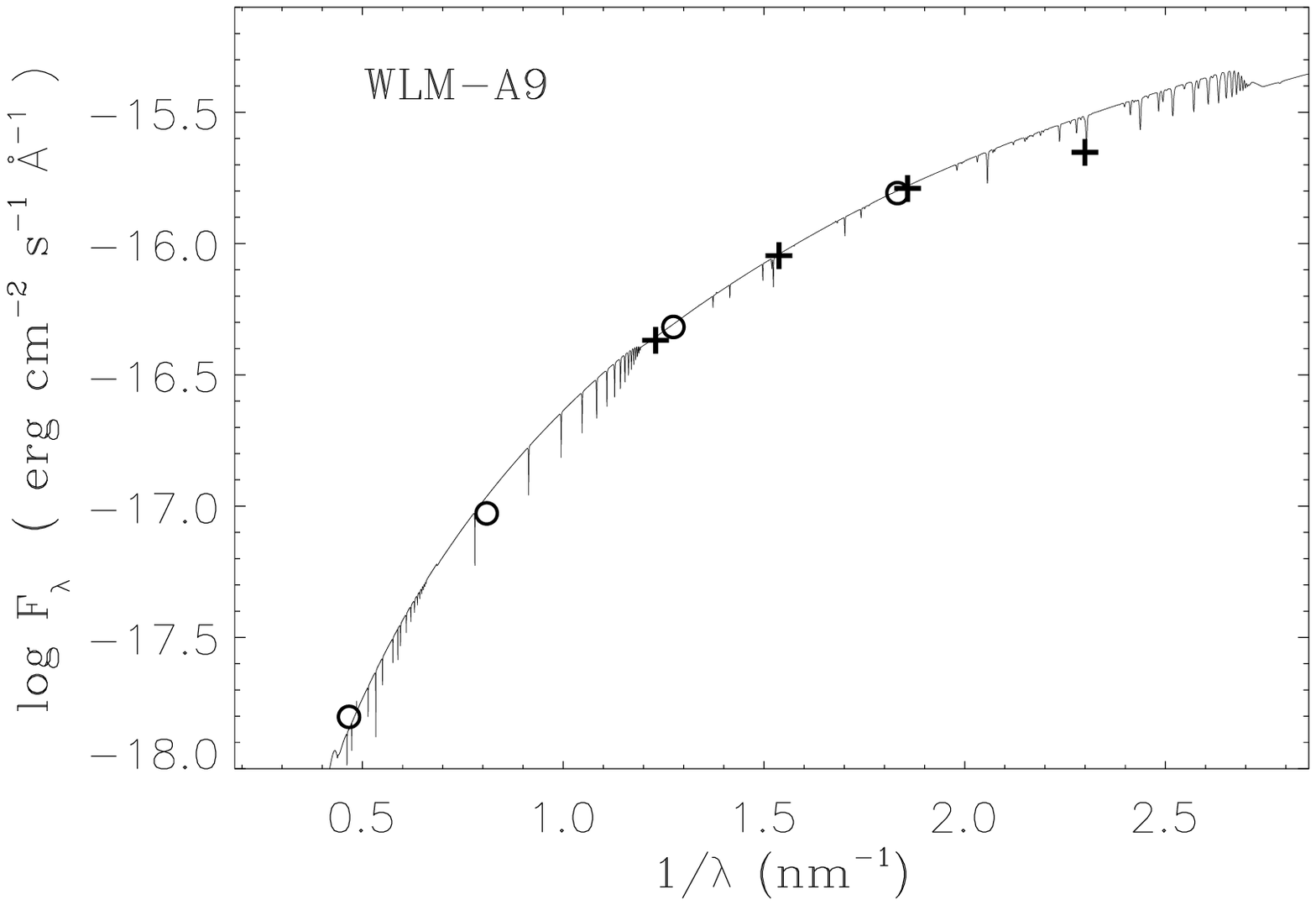}
  \includegraphics[scale=.45]{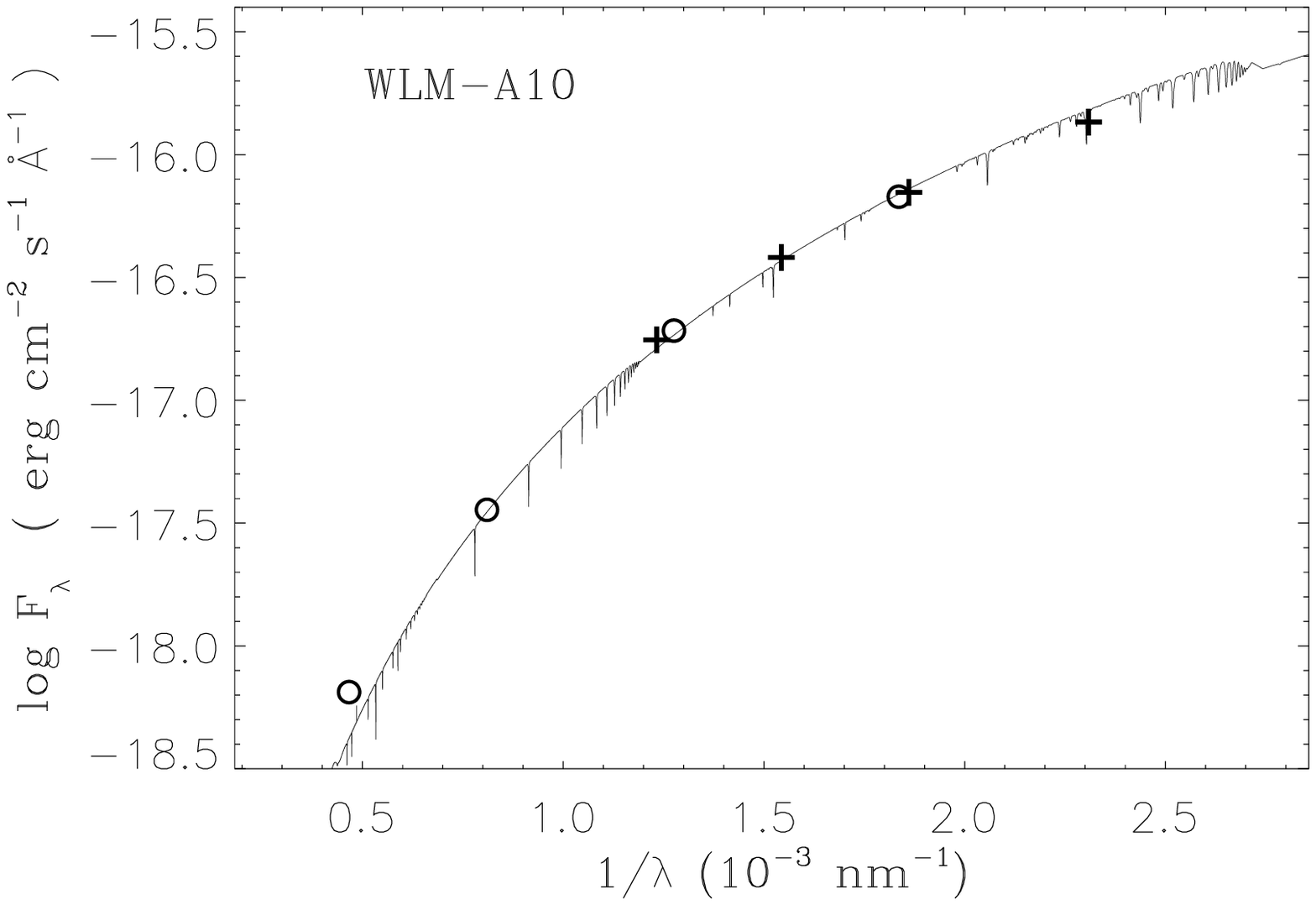}  
  \caption{Synthetic reddened SEDs and photometric measurements for different stars in our WLM combined sample. Circles 
  correspond to V and I photometric values from \citet{bresolin2006} as well as J and Ks photometry from 
  \citet{gieren2008}, and crosses stand for \citet{massey2007} B,V,R and I-band data. \label{fig_seds_ir}}
  \end{center}
\end{figure}

\clearpage
\begin{figure}
  \begin{center}
  \includegraphics[width=0.90\textwidth]{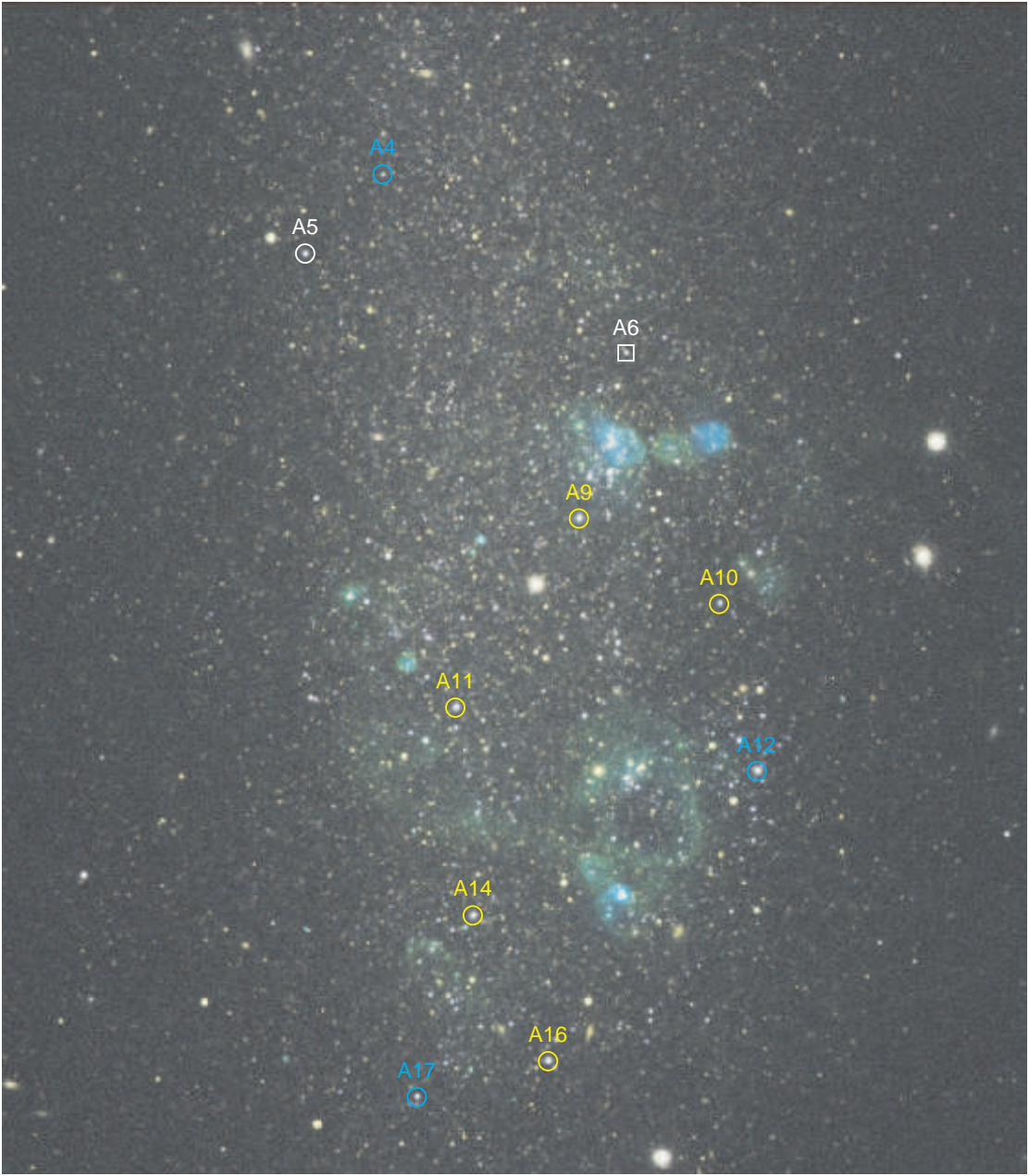}
  \caption{Spatial distribution of the objects. The image is a false color composition using [\ion{S}{2}] (red
  channel), H$\alpha$ (green channel) and [\ion{O}{3}] (blue channel) narrow band images from the Local Group
  Survey \citep{massey2007}. Star's color coding: white circles E(B-V)$<$0.05, cyan 0.05$\le$E(B-V)$<$0.10, 
  yellow 0.10$\le$E(B-V)$<$0.15, and white boxes E(B-V)$\geq$0.15 mag. North is up and East is to the left of the
  image. \label{fig_ebv_halpha}}
  \end{center}
\end{figure}

\clearpage
\begin{figure}
  \begin{center}
  \includegraphics[width=0.90\textwidth]{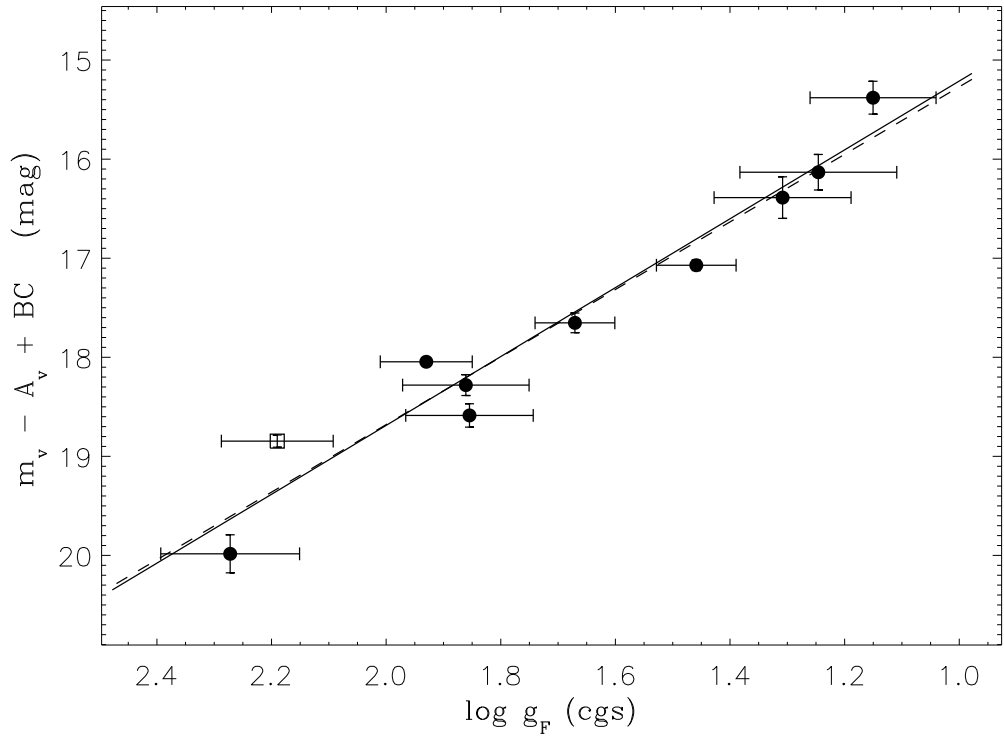}
  \caption{WLM \fglr based on apparent bolometric magnitudes,  
   $\displaystyle m_\mathrm{bol}\,=\,m_\mathrm{v}\,-\,A_\mathrm{v}\,+\,BC$. The solid line is 
  the best linear fit to the data, while the dashed line shows a linear fit when the slope is fixed
  to K08's \fglr based on stars in 8 different galaxies (see text). The star A6 is identified in this
  figure by the square.  \label{fig_fglr_mbol}}
  \end{center}
\end{figure}

\clearpage
\begin{figure}
  \begin{center}
  \includegraphics[width=0.90\textwidth]{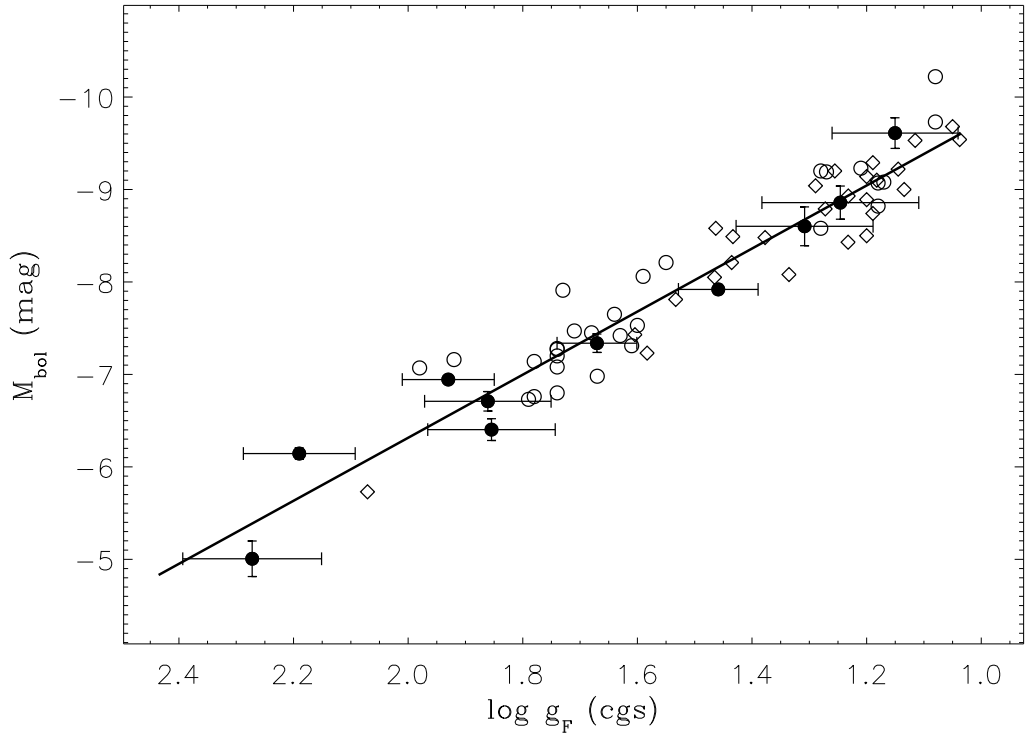}
  \caption{WLM {\sc fglr}. This figure presents the bolometric magnitudes of our WLM stars, once corrected 
  for the distance modulus derived through the {\sc fglr}. The sample of K08 is represented by open symbols 
  (circles: NGC\,300 stars; diamonds: other galaxies. See K08 for details). The thick line corresponds to 
  the \fglr calibration used to determine the distance to WLM.\label{fig_fglr_final}}
  \end{center}
\end{figure}

\clearpage
\begin{figure}
  \begin{center}
  \includegraphics[width=0.90\textwidth]{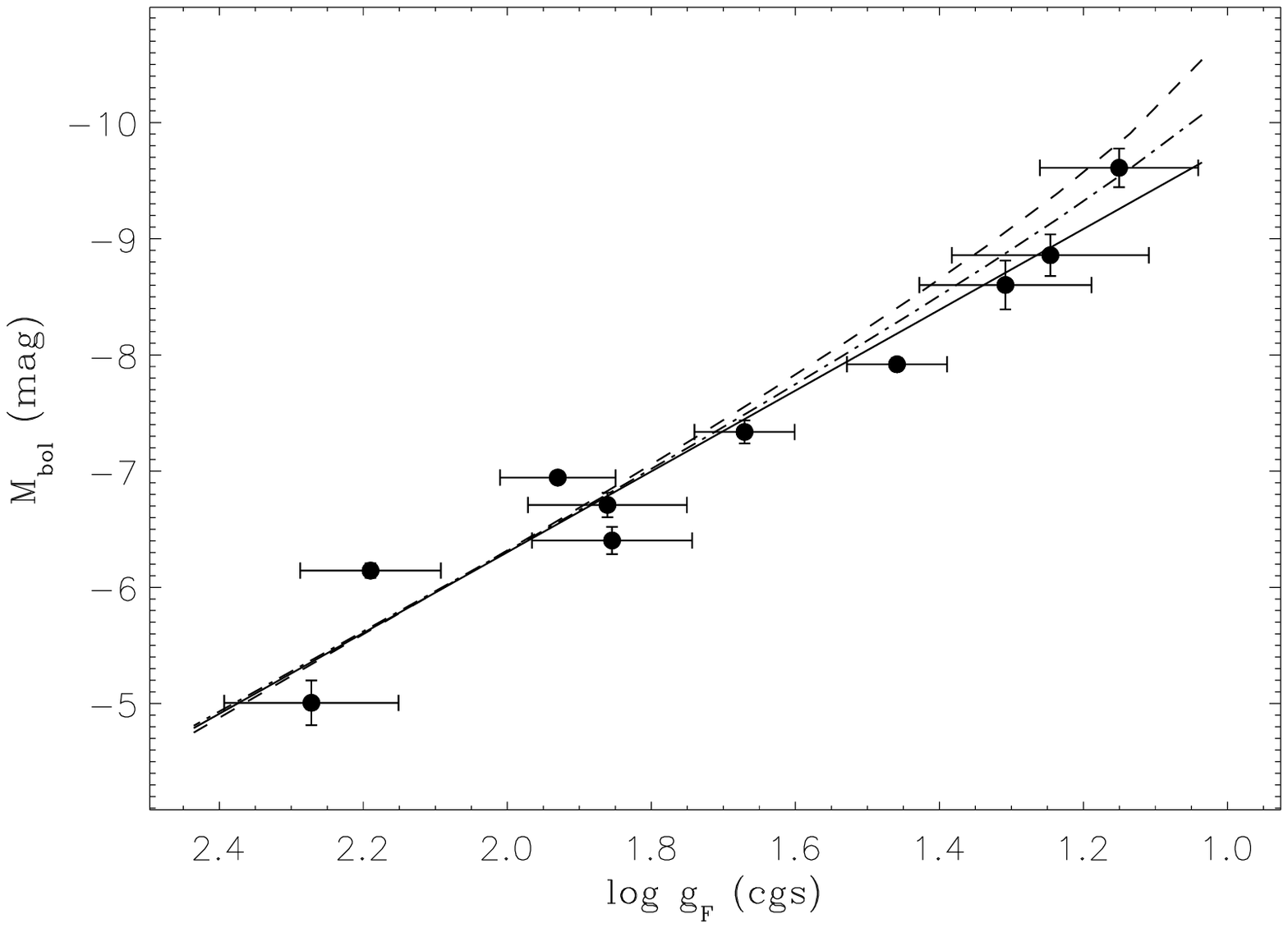}
  \caption{WLM {\sc fglr}. Comparison of empirical and theoretical relationships. The theoretical {\sc fglr}s 
  are obtained from evolutionary models with rotation for SMC metallicity 
  \citep[][dashed line]{maeder2001, meynet2005} and solar metallicity
  \citep[][dot-dash line]{meynet2003}. See text for a detailed discussion.\label{fig_teofglr}}
  \end{center}
\end{figure}

\end{document}